\documentclass[aps,showpacs,twocolumn,superscriptaddress]{revtex4}

\newcommand{\bea}{\begin{eqnarray}}
\newcommand{\eea}{\end{eqnarray}}
\newcommand{\beq}{\begin{equation}}
\newcommand{\eeq}{\end{equation}}

\newcommand{\qz}{\ \stackrel{q \to 0}{\longrightarrow}\ }

\usepackage{amsmath}
\usepackage{amssymb}
\usepackage{latexsym}
\usepackage{graphicx}

\usepackage{graphics}
\usepackage{psfig}
\usepackage{epsfig}
\usepackage{color}
\usepackage{changebar}

\begin{document}


\title{Late Time Analysis for Maximal Slicing of Reissner-Nordstr\"om Puncture Evolutions}


\author{Bernd Reimann}
\affiliation{Max Planck Institut f\"ur Gravitationsphysik,
Albert Einstein Institut, Am M\"uhlenberg 1, 14476 Golm, Germany}
\affiliation{Instituto de Ciencias Nucleares, Universidad Nacional Aut{\'o}noma de M{\'e}xico, A.P. 70-543, M{\'e}xico D.F. 04510, M{\'e}xico}

\author{Bernd Br\"ugmann}
\affiliation{Center for Gravitational Physics and Geometry, Penn State
University, University Park, PA 16802, U.S.A.}


\date{June 10, 2004}


\begin{abstract}
We perform an analytic late time analysis for maximal slicing of the Reissner-Nordstr\"om black hole spacetime.
In particular, we discuss the collapse of the lapse in terms of its late time behavior at the throat and at the event horizon for the even and the puncture lapse. 
In the latter case we also determine the value of the lapse at the puncture.
Furthermore, in the limit of late times slice stretching effects are studied as they arise for maximal slicing of puncture evolutions.
We perform numerical experiments for a Schwarzschild black hole with puncture
lapse and find agreement with the analytical results.
\end{abstract}


\pacs{
04.20.Cv,   
04.25.Dm,   
04.70.Bw,   
95.30.Sf    
\hspace{1mm} Preprint numbers: AEI-2003-107, CGPG-03/12-1
}


\maketitle


\section{Introduction}
\label{sec:introduction}
Choosing a lapse function that results in maximal slicing is a popular coordinate choice when decomposing the 4-dimensional Einstein equations into (3+1)-dimensional form.
This coordinate choice corresponds to the condition that the mean extrinsic curvature of the slices vanishes at all times, leading to a maximal volume property of the hypersurfaces \cite{York79}.

For a single Schwarzschild black hole, maximal slices can be constructed analytically \cite{Estabrook73,Beig98}. 
One can obtain spacelike hypersurfaces which extend from the ``right-hand'' spatial infinity to the ``left-hand'' inner infinity of the extended Schwarzschild spacetime.
These slices give a complete foliation outside the event horizon while approaching a limiting slice inside that does not reach the physical singularity. 
In this sense maximal slices avoid the physical singularity, and even in more general situations maximal slices have been found to be singularity avoiding \cite{Eardley79}.

For these reasons, maximal slicing has been used frequently in numerical simulations of one black hole, e.g.\ \cite{Estabrook73,Bernstein89,Anninos95}, and also in binary black hole mergers in full 3D, e.g.\ \cite{Bruegmann97,Alcubierre00b}.
Furthermore, there now are shift conditions \cite{Alcubierre02a,Bruegmann03} which in many cases overcome the so-called slice stretching problem that previously limited black hole evolutions like those in \cite{Alcubierre00b} to short evolution times, although we do not consider such shift conditions here.

In a previous paper \cite{mypaper1} we established the connection between the analytically known solutions for maximal slicing of the Schwarzschild spacetime \cite{Estabrook73,Beig98} and the maximal slicing computed in the puncture evolution method for black holes \cite{Bruegmann97}.  
Only ``odd'' and ``even'' boundary conditions for the lapse had been considered before, but the numerical lapse of the puncture evolution is not of this type and in the past it was not clear whether the numerically obtained maximal slices exist analytically.
We proved by explicit construction that there exists a maximal slicing of the Schwarzschild spacetime such that the lapse has zero gradient at the puncture, which is the boundary condition that has been observed to hold in numerical evolutions.
We referred to this lapse as puncture lapse or ``zgp'' lapse and presented our results for the non-extremal Reissner-Nordstr\"om metric, generalizing previous constructions of maximal slices.

However, the analytic expressions we found in \cite{mypaper1} are not very suitable for a discussion of the late time behavior of the maximal slices.
In the present paper, in order to provide information about the 4-metric in the limit of late times, we perform a late time analysis (LTA) for maximal slicing with even or puncture lapse.
This study is based on the results in \cite{mythesis,mypaper1} using a method pointed out by \hbox{R.\ Beig} and \hbox{N.\ \'{O} Murchadha} \cite{Beig98}.
In the latter reference the late time behavior of the even lapse at the throat was derived analytically for the Schwarzschild metric. 
Showing that the lapse collapses exponentially there on a time scale given by \hbox{$^{0} \Omega = \frac{3}{4}\sqrt{6}M \approx 1.8371M$}, their analytical result confirmed earlier numerical results by \hbox{L.\ Smarr} and \hbox{J.W.\ York} \cite{Smarr78b}. 

Our LTA will extend \cite{Beig98} by introducing electrical charge to cover the Reissner-Nordstr\"om spacetime. 
Furthermore, the analysis will be extended to four locations of interest in numerical relativity, namely ``from left to right'' - to be understood in terms of a Carter-Penrose diagram or the numerical grid (c.f.\ Fig.~4 or Fig.~5 in \cite{mypaper1}) - the puncture, the left-hand event horizon, the throat, and the right-hand event horizon. 
Studying the behavior of even and puncture lapse at those ``markers'' 
is a useful first step in the discussion of the ``collapse of the lapse''. 

In addition, we will study the \hbox{3-metric} in the limit of late times.
Here, by deriving the late time behavior of the event horizon in volume radius coordinates, we are able to show that a conjecture made in \cite{Beig98} for even boundary conditions does hold there.
Furthermore, for isotropic grid coordinates used in puncture evolutions with vanishing shift, we determine slice stretching effects \footnote{
Since its effects, namely the in-fall of the numerical grid into the black hole and the development of a sharp peak in the metric functions, are a property of the slicing quite independent of the existence of a numerical grid, the name ``grid stretching'' used in some older references is misleading.}.
Whereas those effects frequently have been observed numerically \cite{Anninos95,Anninos95_2}, to our knowledge they are investigated here on an analytic level for the first time.
We discuss the development of a rapidly growing peak in the radial component of the metric (``slice wrapping'') and the ``outward''-drifting of coordinate locations such as the right-hand event horizon (``slice sucking'').

The paper is organized as follows.
In Sec.~\ref{sec:SummarizeMSRN}, we recall the analytic solution for maximal slicing of the Reissner-Nordstr\"om spacetime and discuss briefly odd, even, and ``zgp'' boundary conditions.
In Sec.~\ref{sec:Lapse} we derive the behavior of the even and puncture lapse at late times analytically and compare with numerical results.
In Sec.~\ref{sec:VanShift} slice sucking taking place at the event horizon is studied for volume radius coordinates. 
In addition, for isotropic grid coordinates we discuss the behavior of the \hbox{3-metric} in the limit of late times in order to point out slice stretching effects.
Again we compare with numerical results.
We conclude in Sec.~\ref{sec:conclusion}.
The expansions underlying our LTA are proven in the Appendix.

\section{Maximal Slicing of the Reissner-Nordstr\"om Spacetime}
\label{sec:SummarizeMSRN}

\subsection{General solution in radial gauge}
In this Section we briefly summarize the formulas stated in \cite{mypaper1} for maximal slicing of the Reissner-Nordstr\"om spacetime. 
Following Estabrook et al.\ \cite{Estabrook73} we use the radial gauge,
\beq
\label{eq:4mradial}
	ds^{2} = (-\alpha^{2}+\frac{\beta ^{2}}{\gamma})\:d\tau^{2} 
               + 2\beta\:d\tau dr 
               + \gamma\:dr^2 
               + r^{2}\:d\Omega^{2},
\eeq
to solve the Einstein-Maxwell equations in spherical symmetry under the condition of maximal slicing being vanishing trace of the extrinsic curvature, \hbox{$K = \gamma_{ij} K^{ij} \equiv 0$}. 
We showed that the general solution of the lapse satisfying the corresponding elliptic maximality condition is given by
\beq
\label{eq:generalalpha}
	\alpha(\tau,r) = \frac{\sqrt{p_{C}(\tau,r)}}{r^{2}}
                    \left( D(\tau) + \frac{dC}{d\tau} 
                                     \int^{\infty}_{r} 
                                     \frac{y^{4}\:dy}{p_{C}(\tau,y)^{\frac{3}{2}}}
                    \right).
\eeq
Here $C$ and $D$ are functions of $\tau$ only and the polynomial 
\beq
\label{eq:polynomialproof}
	p_{C}(\tau,r) = r^{4}f(r) + C^{2}(\tau) = r^{4} - 2M r^{3} + qM^{2} r^{2} + C^{2}(\tau)
\eeq
based on
\beq
	f(r) = 1 - \frac{2M}{r} + \frac{qM^{2}}{r^{2}}
\eeq
has been introduced.
As in \cite{mypaper1} we consider the non-extremal Reissner-Nordstr\"om spacetime.
So \hbox{$M > 0$} denotes the mass of the black hole and the dimensionless charge parameter defined by \hbox{$q = \frac{Q^{2}}{M^{2}}$} is restricted to \hbox{$0 \leq q < 1$}.  

Imposing boundary conditions on the lapse fixes the so far undetermined functions $C$ and $D$.
However, independently of these boundary conditions the shift and the radial component of the 3-metric can be expressed as
\beq
\label{eq:betafinal}
        \beta (\tau,r) = \frac{\alpha(\tau,r)\gamma(\tau,r)}{r^{2}} C(\tau)
\eeq
and 
\beq 
\label{eq:gammafinal}
	\gamma (\tau,r) = \frac{1}{f(r) + \frac{C^{2}(\tau)}{r^{4}}} = \frac{r^{4}}{p_{C}(\tau,r)},
\eeq
respectively.

Next we want to re-derive the odd and even lapse and discuss a numerically motivated lapse with zero gradient at the puncture, the ``zgp'' or puncture lapse.

\subsection{Odd Lapse}
\label{subsec:oddlapse}
The first particular lapse choice is referred to as odd (in some references also as antisymmetric) because of its underlying antisymmetry with respect to the throat. 
It corresponds to the boundary conditions
\beq
\label{eq:BCodd}
	\lim_{r \to \infty} \alpha^{\pm}_{odd} = \pm 1 \ \ \forall \tau_{odd}.
\eeq
To measure proper time at infinity (on the right-hand side of the throat denoted by a superscript ``$+$'') the lapse is unity there and due to the antisymmetry the lapse is minus one at the puncture (i.e.\ at the compactified infinity on the left-hand side denoted by ``$-$'').
The odd lapse,
\beq
\label{eq:alphaoddfinal}
	\alpha^{\pm}_{odd}(C,r) = \pm \sqrt{f(r) + \frac{C^{2}}{r^{4}}} 
                                = \pm \frac{\sqrt{p_{C}(r)}}{r^{2}},
\eeq
is obtained by setting in (\ref{eq:generalalpha}) \hbox{$D(\tau_{odd}) = \pm 1$} and \hbox{$\frac{dC}{d\tau_{odd}} = 0$}. 
Note that the slice label $C$ can be chosen independently of the time at infinity $\tau_{odd}$.
Furthermore, whereas event and Cauchy horizon with \hbox{$r_{\pm} = (1 \pm \sqrt{1-q})M$} are given by the real roots of $f(r)$ and are independent of $C$, the throat $r_{C}$ is obtained as a real root of $p_{C}(r)$ and therefore depends on the choice of $C$.
In particular, this implies \hbox{$C = 0$} when starting with initial data where the throat coincides with the event horizon. 
In this case we obtain from (\ref{eq:alphaoddfinal}) the so-called Schwarzschild lapse,
\beq
\label{alphastandard}
	\alpha^{\pm}_{odd} (C = 0,r) = \pm \sqrt{f(r)},
\eeq
from (\ref{eq:betafinal}) a vanishing shift, and from (\ref{eq:gammafinal}) the radial metric component \hbox{$\gamma(r) = \frac{1}{f(r)}$}. 
Hence the static Reissner-Nordstr\"om spacetime written in Schwarzschild coordinates,
\beq
\label{eq:RNradial}
	ds^{2} = -f(r)\:dt^{2} + \frac{1}{f(r)}\:dr^{2} + r^{2}d\Omega^{2},
\eeq
is recovered.

Applying for \hbox{$r\geq r_{+}$} a purely spatial coordinate transformation, 
\bea
\label{eq:rofR}
     r(R) & = & R \left[
                  \left(1+\frac{M}{2R}\right)^{2} 
                  - \frac{qM^{2}}{4R^{2}} 
                  \right] \nonumber \\
	  & = & R + M + \frac{(1-q)M^{2}}{4R},
\eea
chosen such that $r$ and $R$ coincide at infinity and $r_{+}$ is mapped to \hbox{$R_{+} = \frac{1}{2}\sqrt{1-q}M$}, the isotropic lapse 
\beq
\label{eq:alphaisotropic}
       \alpha^{\pm}_{odd}(C=0,R) = \frac{(2R+M)(2R-M) + qM^{2}}{(2R+M)^2 - qM^{2}}
\eeq
is found.
The shift, of course, still vanishes and the \hbox{3-metric} is given by \hbox{$^{(3)}ds^{2} = \Psi^{4}(R) (dR^{2} + R^{2}\:d\Omega^{2})$} with conformal factor
\beq
\label{eq:conformalgamma}
   \Psi^{4}(R) = \left[
		  \left(1+\frac{M}{2R}\right)^{2} 
                  - \frac{qM^{2}}{4R^{2}} 
                 \right]^{2}.
\eeq
One can readily verify that for isotropic coordinates the isometry
\beq 
\label{eq:isometry}
	R \longleftrightarrow \frac{(1-q)M^{2}}{4R} 
\eeq
is present.
In particular, \hbox{$R = 0$} is simply a compactified image of infinity which is referred to as puncture.

\subsection{Even Lapse}
\label{subsec:evenlapse}
Our next example is the even (in some references also called symmetric) lapse satisfying the boundary conditions
\beq
\label{eq:BCeven}
	\lim_{r \to \infty} \alpha^{\pm}_{even} = 1 \ \ \forall \tau_{even}.
\eeq
Here in (\ref{eq:generalalpha}) \hbox{$D(\tau_{even})$} is fixed to be unity and as pointed out in \cite{Estabrook73} the time-dependence manifest in \hbox{$C(\tau_{even})$} can be determined by imposing the requirement of smoothness across the Einstein-Rosen bridge \cite{Einstein35} when passing to the line element (\ref{eq:RNradial}) of the Reissner-Nordstr\"om spacetime in Schwarzschild coordinates. 

In \cite{mypaper1} we have derived expressions for the even lapse and the corresponding ``height function'' which generalize the formulas stated for the Schwarzschild case in \cite{Beig98} by the electric charge.
It turns out that the even lapse can be written as 
\bea
\label{eq:alphaevenfinal}
          \alpha^{\pm}_{even}(C,r) & = & \frac{\sqrt{p_{C}(r)}}{r^{2}}  
			 \frac{dC}{d\tau_{even}} 
			 \frac{\partial t_{even}}{\partial C} \\ \nonumber
                                 \ & = & 
                         - \frac{1}{K_{C}(\infty)}
                           \left( \frac{1}{r - \frac{3}{2}M + \frac{qM^{2}}{2r}} \right. \\ \nonumber
                                 \ & \ & \ \ \ 
                           \left. - \frac{\sqrt{p_{C}(r)}}{r^{2}} K_{C}(r) \right), 
\eea
where the integral
\bea
\label{eq:Kintegral}
    K_{C}(r) =\int\limits^{r}_{r_{C}} \frac{y(y-3M+\frac{3qM^{2}}{2y})\:dy}
		                            {(y - \frac{3}{2}M + \frac{qM^{2}}{2y})^{2} \sqrt{p_{C}(y)}}
\eea
is defined for \hbox{$r \geq r_{C}$}. 
Here the even height function is given by
\beq
\label{eq:tevenfinal}
	t_{even}(C,r) = H_{C}(r)
\eeq
with the integral 
\beq
\label{eq:Hintegral}    
    H_{C}(r) =-\int\limits^{r}_{r_{C}} \frac{C\:dy}{f(y)\sqrt{p_{C}(y)}}
\eeq
again defined for \hbox{$r \geq r_{C}$}.
Note that in (\ref{eq:Hintegral}) the integration across the pole at $r_{+}$ is taken in the sense of the principal value and the corresponding slices extend smoothly through the event horizon $r_{+}$ and the throat $r_{C}$. 
Since proper time is measured at spatial infinity, from (\ref{eq:tevenfinal}) in the limit \hbox{$r \to \infty$} one can infer
\beq
\label{eq:tauevenfinal}
	\tau_{even} (C) = \lim_{r \to \infty} t_{even}(C,r) = H_{C}(\infty).
\eeq

In addition, we want to mention that in deriving these expressions (see \cite{mypaper1} for details) the statement
\bea
\label{eq:dCdH}
	\frac{\partial}{\partial C} H_{C}(r)  = \frac{r^{2}}
                                               {2(r - \frac{3}{2}M + \frac{qM^{2}}{2r})
                                                \sqrt{p_{C}(r)}}
                                          - \frac{1}{2} K_{C}(r) \ \ 
\eea 
has been used, which in the limit \hbox{$r \to \infty$} yields the equation
\beq
\label{eq:HisKhalf}
	\frac{d}{dC} H_{C}(\infty) = - \frac{1}{2} K_{C}(\infty)
\eeq
to be used later on. 
Furthermore, we want to point out that as $\tau_{even}$ is running from zero to infinity, the slice label $C$ takes values ranging from zero to $C_{lim}$.
As shown in \cite{Geyer95} the latter is determined by \hbox{$p_{C}(r_{C}) = 0$} for 
\beq
	r_{C_{lim}} = \frac{1}{4}(3+\sqrt{9-8q})M \qz \frac{3M}{2}
\eeq
to be given by
\bea
\label{eq:defClim}
	C_{lim} & = & \frac{\sqrt{2}}{8} \sqrt{27-36q+8q^{2} + (9-8q)^{3/2}}M^{2} \nonumber \\
                & \qz & \frac{3}{4}\sqrt{3}M^{2} \approx 1.2990 M^{2}.
\eea
With \hbox{$r_{C_{lim}} > 0$} the throat $r_{C}$ - being the ``innermost'' point on a slice labeled by $C$ - never reaches the singularity at \hbox{$r=0$}.
Hence the singularity avoidance of maximal slices for the Reissner-Nordstr\"om spacetime becomes apparent, c.f.\ corollary 3.7 of \cite{Eardley79}.

\subsection{Lapse with zero gradient at the puncture}
\label{subsec:zgplapse}

Next we want to discuss a numerically motivated boundary condition arising for puncture data.
Here a ``radial grid coordinate'' $x$ measures the Euclidean distance from the puncture situated at the origin of a spatial grid and \hbox{$r = x$} for \hbox{$\{r \to \infty,x \to \infty\}$} and \hbox{$r \propto \frac{1}{x}$} for \hbox{$\{r \to \infty, x \to 0\}$} hold. 
On the initial slice this grid coordinate is identical to the isotropic radial coordinate $R$ as in (\ref{eq:rofR}), but $x$ differs at later times depending on the boundary conditions and the shift used in the numerical simulation.  
Demanding as before unit lapse at right-hand infinity,
\beq
\label{eq:BCzgp1}
	\lim_{x \to \infty} \alpha^{+}_{zgp} = \lim_{r \to \infty} \alpha^{+}_{zgp} = 1 \ \ \forall \tau_{zgp},
\eeq
one can label the slices with proper time. 
Numerically, the inner boundary condition arises as a special solution to the maximal slicing equation in the presence of a coordinate singularity in the metric. 
No boundary is actively enforced, but a numerical algorithm finds a solution with sufficiently fast fall-off in the derivative of the lapse such that, even though there is a coordinate singularity in the metric, the lapse is regular there. 
In particular, for the lapse a vanishing gradient at the puncture is obtained which can be formulated as 
\beq
\label{eq:BCzgp2}
	\lim_{x \to 0} \partial_{x} \alpha^{-}_{zgp} \propto \lim_{r \to \infty} r^{2} \partial_{r} \alpha^{-}_{zgp} = 0 
        \ \ \forall \tau_{zgp}
\eeq
making use of \hbox{$\partial_{x} \propto  r^{2} \partial_{r}$} in this limit.

Since the maximality condition determining the lapse is linear in $\alpha$, one can derive as in \cite{mypaper1} the puncture lapse as a superposition,
\beq
\label{eq:linearcombi}
	\alpha^{\pm}_{zgp}(C,r) = \Phi(C) \cdot \alpha^{\pm}_{even}(C,r) 
                                  + (1-\Phi(C)) \cdot \alpha^{\pm}_{odd}(C,r),
\eeq
of the linearly independent odd and even lapse.
Imposing the boundary condition (\ref{eq:BCzgp2}), the ``multiplicator function'' $\Phi$ present in (\ref{eq:linearcombi}) is given by
\beq
\label{eq:PhiofC}
	\Phi(C) = \frac{1}{2}\:\frac{K_{C}(\infty)}{K_{C}(\infty)+\frac{1}{M}}.
\eeq
As we will show later, the integral $K_{C}(\infty)$ diverges in the limit of late times and $\Phi$ hence approaches the \hbox{value $\frac{1}{2}$}.
For this reason one can think of the ``zgp'' lapse as being essentially an average of the odd and even lapse.

Combining the results (\ref{eq:alphaoddfinal}), (\ref{eq:alphaevenfinal}), and (\ref{eq:PhiofC}), we can express the ``zgp'' lapse as 
\bea
\label{eq:alphazgpfinal}
  			\alpha^{\pm}_{zgp}(C,r) 
					       & = &
                         \frac{\sqrt{p_{C}(r)}}{r^{2}}  
			 \frac{dC}{d\tau_{zgp}} 
			 \frac{\partial t_{zgp}}{\partial C} \\
                                            \  & = &
                           - \frac{1}{2} \frac{1}{K_{C}(\infty) +\frac{1}{M}}
                           \left( \frac{1}{r - \frac{3}{2}M + \frac{qM^{2}}{2r}} \right. \nonumber \\
                                            \  & \ & \ \ \ 
                           \left. - \frac{\sqrt{p_{C}(r)}}{r^{2}} 
                                       (K_{C}(r) \pm (K_{C}(\infty) + \frac{2}{M})) \right). \nonumber
\eea
As pointed out in \cite{mypaper1}, the time-dependence is consistently derived from
\bea
\label{eq:zgpdtdC}
	\frac{\partial}{\partial C} t_{zgp}^{\pm}(C,r) 
			  & = & \frac{r^{2}}
                                     {2(r - \frac{3}{2}M + \frac{qM^{2}}{2r})
                                       \sqrt{p_{C}(r)}} \\
                        \ & \ & - \frac{1}{2} \left( K_{C}(r) 
                                          \pm \left(K_{C}(\infty) + \frac{2}{M} \right) \right) \nonumber
\eea
by the ``zgp'' height function 
\beq
\label{eq:tzgpfinal}
	t_{zgp}^{\pm}(C,r) = H_{C}(r) \pm (H_{C}(\infty) - \frac{C}{M})
\eeq
when measuring time at right-hand spatial infinity by
\beq
\label{eq:tauzgpfinal}
	\tau_{zgp} (C)  = \lim_{r \to \infty} t_{zgp}^{+}(C,r) = 2 H_{C}(\infty) - \frac{C}{M}.
\eeq
But note that the time measured at the puncture approaches the finite value
\beq
\label{eq:lefttime}
	\lim_{C \to C_{lim}} \lim_{r\to \infty} t_{zgp}^{-} (C,r) = \frac{C_{lim}}{M}.
\eeq

\section{Late time behavior of the even and puncture lapse}
\label{sec:Lapse}

\subsection{Motivation for a late time analysis}
In practice one is interested in statements for the \hbox{4-metric} (and in particular for the lapse profile) at a given time at infinity $\bar{\tau}$.
This time can be chosen without loss of generality to be \hbox{$0 < \bar{\tau} < \infty$} since for \hbox{$\bar{\tau} < 0$} one could redefine the sign of the time coordinate and for \hbox{$\bar{\tau} = 0$} and \hbox{$\bar{\tau} = \infty$} the solutions corresponding to \hbox{$C=0$} and \hbox{$C = C_{lim}$} are known analytically. 
So as a first step one has to invert the relation $\tau(C)$ - stated for even boundary conditions in (\ref{eq:tauevenfinal}) and for ``zgp'' boundary conditions in (\ref{eq:tauzgpfinal}) - in order to find the appropriate slice label $C$ with which our analytic expressions are to be evaluated in a second step.
Both steps, however, are non-trivial since the primitive for the integrals $H_{C}(r)$, (\ref{eq:Hintegral}), and $K_{C}(r)$, (\ref{eq:Kintegral}), can not be written in closed form. 
Even worse, these integrals are numerically ill-conditioned in the sense that they diverge in the ``late time limit'' \hbox{$C \to C_{lim}$}, as the Schwarzschild radius at the throat $r_{C}$ becomes with $r_{C_{lim}}$ a double counting root of the polynomial $p_{C}(r)$, the root of which appears in the denominator of the integrands of $H_{C}(r)$ and $K_{C}(r)$.
An efficient numerical implementation to calculate these integrals following Appendix 3 of \cite{Thornburg93} can be found in \cite{mythesis}, but the algorithm described there breaks down for $C$ being too close to either $0$ or $C_{lim}$.

\begin{figure}[ht]
	\noindent
	\epsfxsize=85mm \epsfysize=55mm \epsfbox{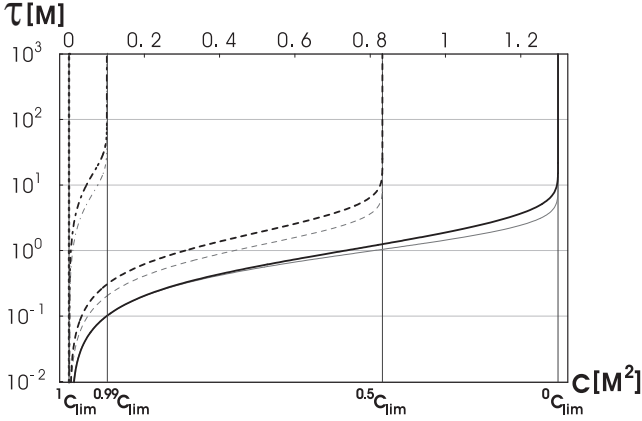}
	\caption{
On a logarithmic scale the time at infinity corresponding to even and ``zero gradient at the puncture'' boundary conditions, $\tau_{even}$ (gray color) and $\tau_{zgp}$ (black), is shown as a function of the slice label C for the four values of the charge parameter \hbox{$q = \frac{Q^{2}}{M^{2}}$} being 0 (solid), 0.5 (dashed), 0.99 (dot-dashed), and 1 (dotted). Note that the range \hbox{$0 \leq C \leq$ $^{q}C_{lim}$} is getting smaller as $^{q}C_{lim}$ declines from its Schwarzschild value \hbox{$^{0}C_{lim} = \frac{3}{4}\sqrt{3}M^{2}$} with increasing $q$ such that \hbox{$^{1}C_{lim} = 0$}. The steep gradient in $\tau(C)$ present as \hbox{$C \to C_{lim}$} causes numerical problems in finding $C$ for a given time at infinity $\bar{\tau}$ as described in the text.}
\label{fig:tauofC}	
\end{figure}

The problem of calculating for a given ``late'' time the appropriate slice label is apparent from Fig.~\ref{fig:tauofC}.
Concentrating in particular on the ``zgp'' Schwarzschild case, we can show (using results of the now following LTA) that the slice labels $^{0}C$ for \hbox{$\bar{\tau}_{zgp} = 10M$}, $100M$ and $1000M$ differ from \hbox{$^{0}C_{lim} = \frac{3}{4}\sqrt{3}M^{2}$} corresponding to \hbox{$\bar{\tau}_{zgp} = \infty$} by only
\bea
	^{0}C_{lim} - {}^{0}C_{zgp} (\bar{\tau}_{zgp} = \ \ \: \: 10M) & < & 9 \cdot 10^{-3}\:\:\:\:M^{2} , \nonumber \\
	^{0}C_{lim} - {}^{0}C_{zgp} (\bar{\tau}_{zgp} = \ \: 100M) & < & 5 \cdot 10^{-24}\:\:M^{2} , \\
	^{0}C_{lim} - {}^{0}C_{zgp} (\bar{\tau}_{zgp} =   1000M) & < & 9 \cdot 10^{-237}M^{2}. \ \ \ \ \ \ \ \nonumber 
\eea
Due to these difficulties when evaluating the analytic expressions, 
hypersurfaces corresponding to \hbox{$\bar{\tau} > 10M$} are rarely shown
in spacetime or embedding diagrams found in the literature for the maximally sliced Reissner-Nordstr\"om or Schwarzschild spacetime \hbox{\cite{Estabrook73,Duncan85,Bernstein89,Thornburg93}}.

\subsection{Concept of a late time analysis}
For practical purposes one is hence interested in approximations for the $C$-dependence of $\tau$. 
The main idea of any LTA now is to introduce an additional time coordinate which is particularly suitable when discussing the limit of late times.

It is interesting to point out that in Eq.~(A3.16a\:-\:c) of \cite{Thornburg93} for the even Schwarzschild case discussed there, an approximately linear relationship 
\beq
\label{eq:Jonathansapprox}
	\tau_{even} \approx  - 0.918555\ln{\left[ \frac{C_{lim}-C}{M^{2}} \right]}M + 0.2865M
\eeq
between the time at infinity $\tau_{even}$ and the ``logarithmic Estabrook-Wahlquist time'' defined by \hbox{$- \ln{\left[ \frac{C_{lim}-C}{M^{2}} \right]}M$} was found empirically.
Anticipating again results of our LTA, this statement turns out to be a very precise approximation for ``late times'' where the constants appearing in (\ref{eq:Jonathansapprox}) can be found analytically to be given by \hbox{$-\frac{1}{2}$$^0\Omega \approx -0.918559M$} and \hbox{$^0\Lambda$$+(\frac{\ln{\sqrt{3}}}{4}-\ln{\left[ M \right]})$$^0\Omega \approx 0.286424M$}.

However, we will not follow up on this logarithmic Estabrook-Wahlquist time further, but instead adopt the approach of \cite{Beig98}, which can be summarized as follows. 
Our LTA is based on $\delta$ introduced by
\beq
\label{eq:defdeltaRN}
	\delta = r_{C} - r_{C_{lim}} 
\eeq
as the difference between the radial coordinate at the throat $r_{C}$ on a slice labeled by $C$ and its final value $r_{C_{lim}}$ corresponding to $C_{lim}$.
Obviously, $\delta$ is lying in \hbox{$0 \leq \delta \leq \sigma_{+}$} when defining
\beq
\label{eq:SIGMA}
	0 \:^{<}_{>} \sigma_{\pm} = r_{\pm}-r_{C_{lim}} 
                         \qz \left\{ \begin{array}{r} \frac{M}{2} \\ -\frac{3M}{2} \end{array} \right. 
\eeq
and $\delta$ approaches zero in the limit of late times.

By expanding the proper time at infinity - which we have found for even and ``zgp'' boundary conditions in (\ref{eq:tauevenfinal}) and (\ref{eq:tauzgpfinal}) - and inverting the relationship, an exponential decay of $\delta$ with $\tau$ is found.
For this decay a ``fundamental time scale'' turns out to be
\bea
\label{eq:Omegais}
    0 < \Omega & = & 
                 - \frac{C_{lim}}{\sqrt{\nu}}
		   \left( 1 - (\frac{\lambda_{+}}{\sigma_{+}} 
		             + \frac{\lambda_{-}}{\sigma_{-}}) \right) \nonumber \\
               & = & \frac{1}{2} \frac{M - \mu}
		             {\sqrt{-\sigma_{+}\sigma_{-}}} \frac{\chi_{-}}{\sqrt{\nu}} \nonumber \\
               & \qz & \frac{3}{4}\sqrt{6}M \approx 1.8371M ,
\eea
where $\Omega$ is defined in terms of the following quantities:
\bea 				
\label{eq:MU}	
	0 < & \mu & = (1+\sqrt{9-8q})M \qz 4M , \\
\label{eq:NU}
	0 < & \nu & = \frac{1}{4} (9-8q + 3\sqrt{9-8q})M^{2} \qz \frac{9}{2}M^{2} , \\
\label{eq:LAMBDA}
	0 \:^{<}_{>} &\lambda_{\pm} & = \left( 1 \pm \frac{2-q}{2\sqrt{1-q}} \right)M 
                     \qz \left\{ \begin{array}{c} 2M \\ 0 \end{array} \right. , \\ 
\label{eq:XI}
	0 \leq &\xi_{\pm} & = \frac{\lambda_{\pm}}{\sigma_{\pm} \sqrt{\sigma_{\pm}^{2}+\sigma_{\pm}\mu+\nu}} \nonumber \\
	               & \ &  \ \ \ \ \ \ln {\left[ \pm \frac{\sigma_{\pm}\mu 
				      + 2(\nu+\sqrt{\nu}\sqrt{\sigma_{\pm}^{2}+\sigma_{\pm}\mu+\nu})}
					                 {\sigma_{\pm}(\mu+2(\sigma_{\pm}
					+\sqrt{\sigma_{\pm}^{2}+\sigma_{\pm}\mu+\nu}))} \right]} \nonumber \\
                       & \qz & \left\{ \begin{array}{c} \frac{8}{3\sqrt{3}} 
                                                     \ln{\left[ \frac{9\sqrt{6} + 22}{5+3\sqrt{3}} \right]} \frac{1}{M} 
                 				     \approx 2.2528\frac{1}{M}
                                                     \\ 0 \end{array} \right.  , \\
\label{eq:CHI}
        0 \:^{\leq}_{>} & \chi_{\pm} & =  - \frac{1}{8} 
			\left(9-2q \pm \frac{9(-3+2q)}{\sqrt{9-8q}}\right) M^{2} \nonumber \\
			& \qz & \left\{ \begin{array}{c} 0 \\ -\frac{9}{4}M^{2} \end{array} \right. .
\eea
Note that a time scale \hbox{$^{0} \Omega \approx 1.82 M$} has been found numerically for the Schwarzschild spacetime in \cite{Smarr78b} and that its analytical value \hbox{$^{0} \Omega = \frac{3}{4}\sqrt{6}M \approx 1.8371M$} has been derived in \cite{Beig98}. 

Then if one is interested in studying the lapse, one has to expand the lapse functions, (\ref{eq:alphaevenfinal}) or (\ref{eq:alphazgpfinal}), in terms of $\delta$ and discuss the limit $\delta \to 0$.
When calculating these expansions the following important expressions arise:
\bea
\label{eq:ETA}
        0 > & \eta & = \frac{1}{\sqrt{\nu}} 
 			   \left(	
			      1 + \chi_{+}\:\frac{4}{(M - \mu)^{2}} 
			        + \chi_{-}\:\frac{3\mu^{2} - 4\nu}{8\nu^{2}} 	
			   \right) \nonumber \\
	            & \ & \ \ \ \ \ \ln{
			      \left[
				\frac{\sigma_{+}\mu + 2(\nu+\sqrt{\nu}
                                        \sqrt{\sigma_{+}^{2} + \sigma_{+}\mu + \nu})}
				     {\sigma_{+}(\mu+2\sqrt{\nu})}
			      \right]} \nonumber \\
                    & \ & + \chi_{+} \frac{16(2\sigma_{-} - M +2\sqrt{\sigma_{+}^{2} + \sigma_{+}\mu + \nu})}
                                        {(2\sigma_{-} - M)^{2}(M-\mu)^{2}} \nonumber \\
                    & \ & + \chi_{-} \frac{3\sigma_{+}^{2}\mu+(2\nu-3\sigma_{+}\mu)
						\sqrt{\sigma_{+}^{2} + \sigma_{+}\mu + \nu}}
					{(2\sigma_{+}\nu)^{2}} \nonumber \\
                    & \qz & \frac{1}{36} 
                            \left( 7\sqrt{2} 
                            \ln{\left[ \frac{9\sqrt{6}+22}{3\sqrt{2}+4} \right]} 
				      -6(3\sqrt{3}+2) \right) \frac{1}{M} \nonumber \\ 
		    & \approx & - 0.7385 \frac{1}{M} ,
\eea
and 
\bea
\label{eq:Lambdais}
            \Lambda & = & - \Omega \ln{\left[ \frac{2\sqrt{\nu} + \mu}{8\nu} \right]} 
                             - C_{lim} (\xi_{+} + \xi_{-}) \nonumber \\
                    & \qz & \:{}^{0}\Omega \ln{\left[ M  \right]} 
			    + (\frac{3}{4}\sqrt{6} \ln{ \left[ 18(3\sqrt{2} - 4) \right]} \nonumber \\
		    & \ & \ \ \ \ \ - 2 \ln{ \left[ \frac{3\sqrt{3}-5}{9\sqrt{6}-22} \right]})M \nonumber \\
		    & \approx & 1.8371M \ln{\left[ M  \right]} - 0.2181M.
\eea
Here $\Lambda$ is playing the role of the constant $A$ as defined in Eq.~(3.41) of \cite{Beig98}. 
As the reader might guess by now, the calculations are rather involved. 
For this reason we have only performed the $\delta$-expansions at the puncture, the left-hand event horizon, the throat and the right-hand event horizon.
But information obtained at these markers essentially allows us to sketch the lapse profile and discuss in leading order the ``collapse of the lapse''.
We have summarized our results for even and ``zgp'' boundary conditions in Subsecs.~\ref{subsec:LTAevenlapse} and \ref{subsec:LTAzgplapse}, respectively.
The proof of those statements can be found in the Appendix.

\subsection{Late time behavior of the even lapse}
\label{subsec:LTAevenlapse}
Expanding proper time at infinity in $\delta$ and inverting the relation, $\delta$ is found to decay exponentially with time as 
\beq
\label{eq:deltaevenRN}
	\delta = \kappa_{even} \exp \left[ -\frac{\tau_{even}}{\Omega_{even}} \right] 
		+ {\cal O} (\exp \left[ - 2 \frac{\tau_{even}}{\Omega_{even}} \right]).
\eeq	
We would like to refer to the time scale
\beq 
\label{eq:OmegaevenRNis}
	\Omega_{even} = \Omega \qz \frac{3}{4}\sqrt{6}M \approx 1.8371M
\eeq
defined in (\ref{eq:Omegais}) as being the fundamental time scale which characterizes the collapse of the lapse when maximally slicing a Reissner-Nordstr\"om black hole.
The pre-exponential factor in (\ref{eq:deltaevenRN}) turns out to be 
\beq
\label{eq:kappaevenis}
	\kappa_{even} =  \exp \left[ \frac{\Lambda}{\Omega} \right]
                      \qz \exp \left[ \frac{{}^{0}\Lambda}{{}^{0}\Omega} \right] \approx 0.8880M.
\eeq
Due to its boundary conditions the even lapse is one at both puncture and infinity.
Expanding the even lapse in $\delta$ at the throat, which is the center of symmetry, yields
\bea
\label{eq:lapsethroatRN}
		\alpha^{\pm}_{even}\mid_{r_{C}} & = & 
			\frac{2 r_{C_{lim}}\sqrt{\nu}}
                     	{\chi_{-}(M - \mu)}\:\delta 
                	+  {\cal O} (\delta^{2}) \nonumber \\
        	& \qz & \frac{2\sqrt{2}}{3} \frac{\delta}{M} 
                	+  {\cal O} (\delta^{2}) \nonumber \\
        	& \approx & 0.9428 \frac{\delta}{M} +  {\cal O} (\delta^{2}). 
\eea
We want to mention here that for the Schwarzschild spacetime this decay of ${\cal O}(\delta)$ at the throat has been derived already in \cite{Beig98}.
Extending the latter study we obtain for the lapse at the event horizon the expansion
\bea
\label{eq:evenlapseREHRN}
	      \alpha^{\pm}_{even} \mid _{r_{+}} 
              & = &
	 	\frac{C_{lim}}{r_{+}^{2}}  
	      + \frac{\sqrt{\nu}}{\chi_{-}}
		      \left(
			-\frac{4}{\sigma_{+}-\sigma_{-}} 
			-\frac{1}{r_{+}^{2}}
			 (C_{lim}\eta  \right. \nonumber \\
	      & \ & \ \ \left. 
			- \frac{M - \mu}{2\sqrt{-\sigma_{+}\sigma_{-}}}\frac{\chi_{-}}{\sqrt{\nu}})
		      \right)\:\delta^{2}
	      + {\cal O} (\delta^{3}) \nonumber \\
	      & \qz & 
		\frac{3}{16}\sqrt{3} 
		+ \frac{1}{144\sqrt{2}}
		\bigg( 7\sqrt{6}\ln{\left[ \frac{9\sqrt{6}+22}{3\sqrt{2}+4} \right]} \nonumber \\
	      & \ & \ \ +330 - 12\sqrt{3} - 36\sqrt{6} \bigg) \frac{\delta^{2}}{M^{2}}
		+ {\cal O} (\delta^{3}) \nonumber \\
	      & \approx &
		0.3248 + 1.2265\frac{\delta^{2}}{M^{2}} +  {\cal O} (\delta^{3}),
\eea
which because of symmetry is identical for the left-hand and right-hand event horizon. 
Note that the finite value approached there asymptotically is in agreement with our conjecture made in \cite{mypaper1}.

\subsection{Late time behavior of the puncture lapse}
\label{subsec:LTAzgplapse}
Similarly, the late time behavior of $\delta$ as a function of $\tau_{zgp}$ is determined by
\beq
\label{eq:deltazgpRN}
	\delta = \kappa_{zgp} \exp{\left[ -\frac{\tau_{zgp}}{\Omega_{zgp}} \right]}
		+ {\cal O} (\exp{\left[ - 2 \frac{\tau_{zgp}}{\Omega_{zgp}} \right]}),
\eeq
where the time scale of the exponential decay is found to be twice the fundamental time scale,
\beq
\label{eq:OmegazgpRNis}
	\Omega_{zgp} = 2 \Omega \qz \frac{3}{2}\sqrt{6}M
	             \approx 3.6742M.
\eeq
Here the prefactor $\kappa_{zgp}$ can be calculated as 
\bea
\label{eq:kappazgpis}
	\kappa_{zgp} & = & \exp{\left[ \frac{\Lambda - \frac{1}{2}C_{lim}\frac{1}{M}}{\Omega} \right]} \nonumber \\
                     & \qz & \exp \left[ \frac{{}^{0}\Lambda
					-\frac{1}{2}{}^{0}C_{lim}\frac{1}{M}}{{}^{0}\Omega} \right] 
		     \approx 0.6236M. \ \ 
\eea
For the lapse with zero gradient at the puncture constructed as a superposition of the odd and even lapse, the lapse is again one at infinity in order to measure proper time.
The multiplicator function $\Phi$ appearing in the linear combination (\ref{eq:linearcombi}) can be expanded in $\delta$ as 
\bea
	\Phi & = & \frac{1}{2} - \frac{\sqrt{\nu}}{2\chi_{-}M}\:\delta^{2} +  {\cal O} (\delta^{3}) \nonumber \\
             & \qz & \frac{1}{2} + \frac{\sqrt{2}}{3}\frac{\delta^{2}}{M^{2}} + {\cal O} (\delta^{3}) \nonumber \\
             & \approx & \frac{1}{2} + 0.4714\frac{\delta^{2}}{M^{2}} + {\cal O} (\delta^{3}).
\eea
Therefore, at leading order the puncture lapse at late times is the average of the odd and even lapse.
Next one can derive statements for the puncture lapse from ``left to right'' in the direction of increasing lapse.
It turns out that the ``zgp'' lapse collapses at the puncture,
\bea
\label{eq:zgplapsepunctureRN}
		\alpha^{-}_{zgp}\mid_{x=0} 
                & = & - \frac{\sqrt{\nu}}{\chi_{-}M}\:\delta^{2} +  {\cal O} (\delta^{3}) \nonumber \\
                & \qz & \frac{2\sqrt{2}}{3}\frac{\delta^{2}}{M^{2}} + {\cal O} (\delta^{3}) \nonumber \\
		& \approx & 0.9428\frac{\delta^{2}}{M^{2}} + {\cal O} (\delta^{3}), 
\eea
and at the left-hand event horizon,
\bea
\label{eq:zgplapseLEHRN}		
		\alpha_{zgp}^{-}\mid_{r_{+}} 
		& = &
		\frac{\sqrt{\nu}}{\chi_{-}}		  
		    \left(
			-\frac{2}{\sigma_{+}-\sigma_{-}} \right. \nonumber \\
		& \ & \ \ \left.
			-\frac{C_{lim}}{r_{+}^{2}}(\frac{1}{M} + \frac{\eta}{2})
		    \right)\:\delta^{2} + {\cal O} (\delta^{3}) \nonumber \\
		& \qz &
 		\frac{1}{288\sqrt{2}}
		\bigg(7\sqrt{6} \ln{\left[ \frac{9\sqrt{6}+22}{3\sqrt{2}+4} \right]} \nonumber \\
		& \ & \ \ + 330 + 60\sqrt{3}\bigg)
                \frac{\delta^{2}}{M^{2}} 
		+ {\cal O} (\delta^{3}) \nonumber \\
		& \approx &
		1.1359\frac{\delta^{2}}{M^{2}} + {\cal O} (\delta^{3}),
\eea
in order ${\cal O}(\delta^2)$.
At the throat, however, since the odd lapse vanishes there and does not contribute to the ``late time average'', with 
\bea
\label{eq:zgplapsethroatRN}
		\alpha^{\pm}_{zgp}\mid_{r_{C}} 
                & = & \frac{r_{C_{lim}}\sqrt{\nu}}
                       {\chi_{-}(M - \mu)}\:\delta +  {\cal O} (\delta^{2}) \nonumber \\
	        & \qz & \frac{\sqrt{2}}{3} \frac{\delta}{M} +  {\cal O} (\delta^{2}) \nonumber \\
                & \approx & 0.4714 \frac{\delta}{M} +  {\cal O} (\delta^{2})  
\eea
the prefactor of the leading term of order ${\cal O} (\delta)$ has a value which is half the one found for even boundary conditions.
Finally, for the right-hand event horizon we obtain
\bea
\label{eq:zgplapseREHRN}
		\alpha_{zgp}^{+} \mid _{r_{+}} 
               & = &
		\frac{C_{lim}}{r_{+}^{2}}
	      + \frac{\sqrt{\nu}}{\chi_{-}}
		      \left(
			-\frac{2}{\sigma_{+}-\sigma_{-}}
		        -\frac{1}{2r_{+}^{2}}
                        (C_{lim}\eta \right. \nonumber \\
	       & \ & \ \ \left.
			- \frac{M-\mu}{\sqrt{-\sigma_{-}\sigma_{+}}} \frac{\chi_{-}}{\sqrt{\nu}})
		      \right)\:\delta^{2}
	      + {\cal O} (\delta^{3}) \nonumber	\\
		& \qz &
                \frac{3}{16}\sqrt{3}
                + \frac{1}{288\sqrt{2}}
		\bigg( 7\sqrt{6} \ln{\left[ \frac{9\sqrt{6}+22}{3\sqrt{2}+4} \right]} \nonumber \\
                & \ & \ \ + 330 - 72\sqrt{6} - 12\sqrt{3} \bigg)
                \frac{\delta^{2}}{M^{2}} 
		+ {\cal O} (\delta^{3}) \nonumber \\
		& \approx & 0.3248 + 0.3967\frac{\delta^{2}}{M^{2}} + {\cal O} (\delta^{3}), 
\eea
where again the finite value $\frac{C_{lim}}{r_+^2}$ is approached asymptotically. 
The reader should compare these results with the corresponding ones obtained for even boundary conditions.

\subsection{Analytical and numerical results for the Schwarzschild puncture lapse}
\label{subsec:LTACactus}
The late time expansions for the even and the ``zgp'' lapse have been tested both analytically and numerically.

\begin{figure}[!ht]
	\noindent
	\epsfxsize=82mm \epsfysize=183mm \epsfbox{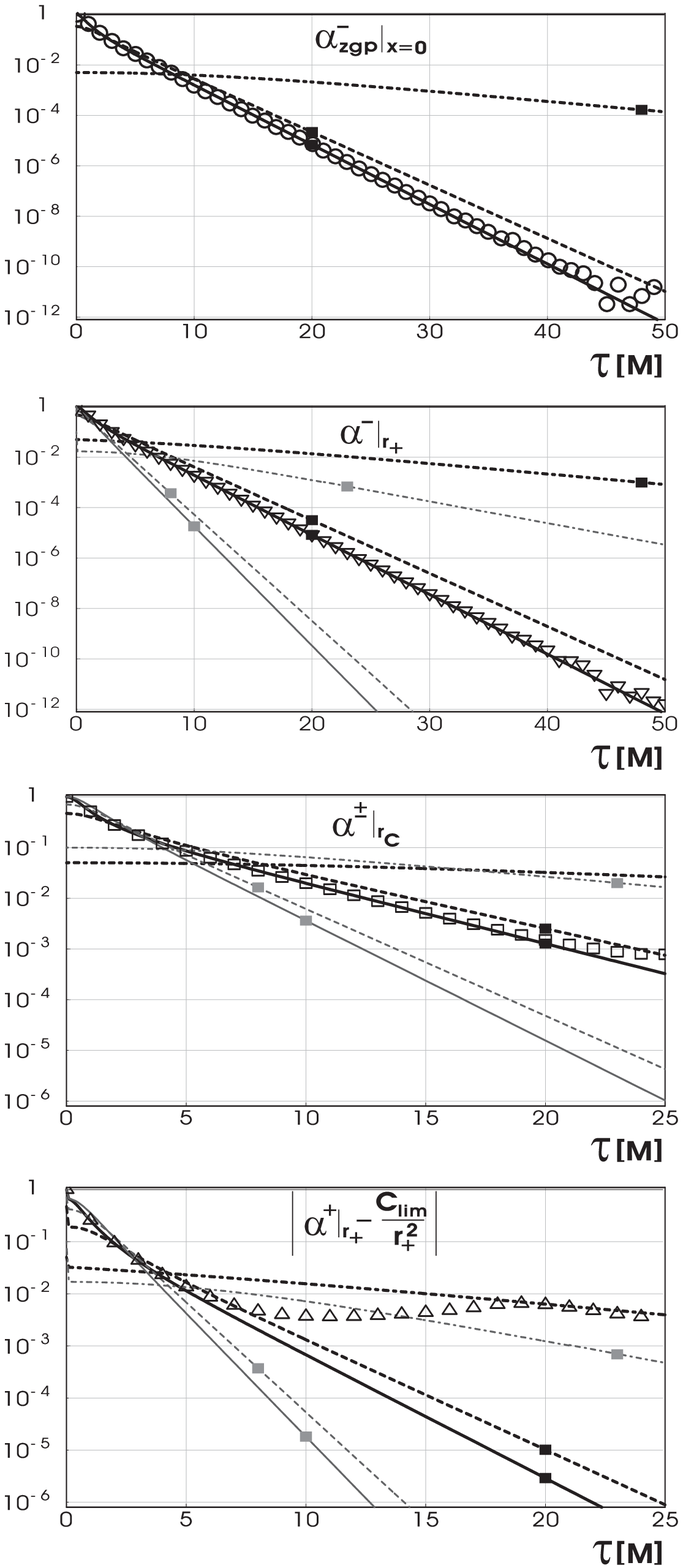}
	\caption{For puncture, left-hand event horizon, throat and right-hand event horizon from top to bottom the value of the lapse as a function of time at infinity is shown. The lines - plotted in gray or black corresponding to even or ``zgp'' boundary conditions - represent analytic results obtained for \hbox{$q = 0$} (solid), $0.5$ (dashed) and $0.99$ (dot-dashed). In addition numerical results obtained for vanishing charge and ``zgp'' boundary conditions are plotted as symbols. Those are found to agree very well with our analytic statements at the puncture (circles) and the left-hand event horizon (downward pointing triangles). The deviations occurring at the throat (boxes) and the right-hand event horizon (upward pointing triangles) are explained in the text.}
	\label{fig:lapsestatements}
\end{figure}

In Fig.~\ref{fig:lapsestatements} analytic results for the lapse are shown corresponding to different values of the charge parameter $q$ and both even and ``zgp'' boundary conditions.
Since the even lapse at the puncture is identically one it has not been plotted.
We want to point out that the analytic solutions $\alpha_{even}^{\pm}$, (\ref{eq:alphaevenfinal}), and $\alpha_{zgp}^{\pm}$, (\ref{eq:alphazgpfinal}), have been evaluated at the appropriate values of $r$ in the manner described in \cite{mythesis} to produce the lines up to the box symbols, where the algorithm breaks down.
Note that these lines go over smoothly when making use of the appropriate late time expansions of Subsecs.~\ref{subsec:LTAevenlapse} and \ref{subsec:LTAzgplapse}, which verifies the leading order terms of our LTA. 

In addition, we have performed numerical simulations with ``zgp'' maximal slicing and vanishing shift.
The particular results presented here were obtained with the Cartoon method described in \cite{Alcubierre99a} and implemented for the present purpose for spherical symmetry in the three-dimensional BAM code \cite{Bruegmann03} (our previous implementations were for axisymmetry). 
The elliptic maximal slicing equation is solved with the BiCGSTAB algorithm, and for the details of the puncture evolution method see \cite{Alcubierre02a}.
There are 512 equidistant points in the radial direction $x$ with a grid spacing of \hbox{$\triangle x = 0.025M$} and the outer boundary at about \hbox{$x = 12.8M$}.

Since the puncture is staggered between grid points, we obtain the value of the lapse at \hbox{$x = 0$} by polynomial interpolation using two grid points on each side.
In order to calculate the lapse by interpolation at the throat and the left-hand and right-hand event horizon, we first have to locate those on the grid as a function of time, c.f.\ Fig.~5 of \cite{mypaper1}. 
For this task we compute the Schwarzschild radial coordinate $r$ from the prefactor of the angular part of the metric as e.g.\ in \cite{Brandt94,mypaper1}, and identify the throat $x_{C}$ on a certain slice as the minimum of $r$ and the left-hand and right-hand event horizon $x_{C}^{\pm}$ as isosurfaces with \hbox{$r = {}^{0}r_{+} =2M$}. 

As one can infer from Fig.~\ref{fig:lapsestatements}, the numerical results agree quite well up to \hbox{$\tau_{zgp} \approx 40M$} with the analytically derived ``zgp'' Schwarzschild lapse and its late time expansions at the puncture and at the left-hand event horizon.
Later on deviations occur since by then both round-off errors and noise introduced by the outer boundary contribute significantly.

The agreement at the throat and at the right-hand event horizon is less convincing, note the different time scale of the plots. 
But as the maximal slices approach at late times the limiting slice \hbox{$r = {}^{0}r_{C_{lim}} = \frac{3M}{2}$} asymptotically, it is very difficult to determine the location $x_{C}$ of the throat as a minimum of the Schwarzschild radial coordinate $r$ since the minimum is very shallow, c.f.\ the lower plot of Fig.~\ref{fig:Metric}.
Furthermore, since the lapse rises quickly from zero to unity near the right-hand event horizon, c.f.\ the upper plot of Fig.~\ref{fig:Metric}, small deviations from the correct $x_{C}^{+}$ result in significant error for the value of the lapse.   
Those errors dominate at later times when the difference between the lapse and its finite asymptotic value is measured, which is only of order ${\cal O}(\delta^2)$ and therefore poses a stringent test.

\section{Slice Stretching for Vanishing Shift}
\label{sec:VanShift}
The foliation with maximal slices - shown in Kruskal-Szekeres and Carter-Penrose diagrams in \cite{mypaper1} for the Schwarzschild spacetime and in \cite{mythesis} for the Reissner-Nordstr\"om spacetime - is uniquely determined by the lapse function.
However, we are still free to choose the shift function and adopt different spatial coordinates on the maximal slices. 
In order to distinguish the 4-metric in those spatial coordinates from the radial gauge, (\ref{eq:4mradial}), we denote in capital letters with $B$ the shift and with $G_{ij}$ and $G$ the \hbox{3-metric} and its radial component, respectively.  

Demanding a vanishing shift throughout the evolution, \hbox{$B \equiv 0$}, is a trivial choice of the shift function which has been used in numerical relativity for its simplicity and which we will study in the following.
In the context of maximal slicing, \hbox{$K \equiv 0$}, this implies that the determinant of the 3-metric is time-independent. 
Hence the singularity avoiding property comes to light as the variation of the local volume remains fixed \cite{Choptuik86}. 

We will use the time-independency of $\begin{rm}{det}\end{rm}\{ G_{mn} \}$ to study in Subsecs.~\ref{subsec:VolumeRadius} and B the volume radius coordinate $\rho$. 
As in \cite{Beig98} we will restrict ourselves to even boundary conditions. 
Here we are able to verify at the event horizon a conjecture regarding the late time behavior of the lapse at ``large'' $\rho$ formulated for Schwarzschild in this reference.

Again in the context of \hbox{$B \equiv 0$}, in Subsecs.~\ref{subsec:Gridcoordinates} to F we look at radial isotropic grid coordinates $x$.
Those have been constructed in \cite{mypaper1} such that the \hbox{4-metric} coincides at all times with output from a numerical evolution of puncture initial data.
Using ``from left to right'' the puncture, the left-hand event horizon, the throat and the right-hand event horizon as markers, we are able to point out slice stretching effects in the limit of late times.

\subsection{Volume radius coordinates}
\label{subsec:VolumeRadius}
Vanishing shift implies Eulerian observers obtained by choosing coordinates $\rho(C(\tau),r)$ such that the level surfaces \hbox{$\rho = \begin{rm}{const}\end{rm}$} are timelike cylinders orthogonal to the slicing. 
As pointed out in \cite{Beig98} for the Schwarzschild spacetime and generalized here in a straightforward way for the Reissner-Nordstr\"om spacetime, one of these timelike cylinders, namely the one given by \hbox{$r = r_{C}$}, is already known. By using the following expression for the volume radius,
\beq
\label{eq:volumerho}
	\rho^{3}(C(\tau_{even}),r) = 3 \int\limits_{r_{C}}^{r} \frac{y^{4}\:dy}{\sqrt{p_{C}(y)}},
\eeq
on each slice such a coordinate is found for even boundary conditions.
To see this one may as a preparatory work verify that the partial derivatives of $\rho$ with respect to $r$ and $C$ are given by
\beq
\label{eq:drhodr}
	\frac{\partial}{\partial r} \rho
              = \frac{1}{\rho^{2}} \frac{\partial}{\partial r} 
                \int\limits_{r_{C}}^{r} \frac{y^{4}\:dy}{\sqrt{p_{C}(y)}}
              = \frac{1}{\rho^{2}} \frac{r^{4}}{\sqrt{p_{C}(r)}}
\eeq
and
\beq
\label{eq:drhodC}
	\frac{\partial}{\partial C} \rho
           =  \frac{1}{\rho^{2}} \frac{\partial}{\partial C} 
                \int\limits_{r_{C}}^{r} \frac{y^{4}\:dy}{\sqrt{p_{C}(y)}}
           =  \frac{C}{\rho^{2}} \frac{\partial}{\partial C} t_{even}(C,r),
\eeq
respectively. 
In deriving (\ref{eq:drhodC}) the reader should remember the notation \hbox{$t_{even} (C,r) = H_{C}(r)$}, (\ref{eq:tevenfinal}). Furthermore, the expression
\beq
\label{eq:dCdHmHisI}
     \frac{\partial}{\partial C}\left( H_{C}(r) - H_{C}(\infty) \right) 
     	 = \int^{\infty}_{r} 
           \frac{y^{4}\:dy}{p_{C}(y)^{\frac{3}{2}}},   
\eeq
the formula
\beq
\label{eq:drCdC}
	\frac{dr_{C}}{dC} = - \frac{C}{2r_{C}^{2}(r_{C} - \frac{3}{2}M + \frac{qM^{2}}{2r_{C}})}
\eeq
obtained by differentiating \hbox{$p_{C}(r_{C}) = 0$} with respect to $C$ and the statement (\ref{eq:dCdH}) are helpful.

Hence for even boundary conditions the Reissner-Nordstr\"om metric can be written as
\beq
\label{eq:evenrhometric}
      ds^{2}   =  -\alpha_{even}^{2}\:d\tau_{even}^{2}  
               + \left( \frac{\rho}{r} \right)^{4}\:d\rho^2 
               + r^{2}\:d\Omega^{2}.			
\eeq
This result stated in \cite{Beig98} one can immediately check remembering the formulas (\ref{eq:alphaevenfinal}), (\ref{eq:betafinal}) and (\ref{eq:tauevenfinal}) for lapse, shift and time at infinity in the even case.
One should observe that in (\ref{eq:evenrhometric}) the determinant of the \hbox{3-metric} is given independently of $r(C(\tau_{even}),x)$ by \hbox{$\det{\left\{ G_{mn} \right\}} = \rho^{4} \sin^{2}{\theta}$} and hence is time-independent. 

\subsection{The location of the event horizon in the limit of late times}
\label{subsec:BeigsCon}
Next we want to show that a conjecture formulated by \hbox{R.\ Beig} and N.~\'{O} Murchadha in \cite{Beig98} holds at the event horizon.
For the Schwarzschild spacetime the conjecture was that the collapse of the even lapse should take place according to 
\bea
\label{eq:Beigsconjecture1}
	^{0}\alpha_{even}(\tau_{even},{}^{0}\rho) 
		& = & B({}^{0}\rho) \exp \left[ -\frac{\tau_{even}}{{}^{0}\Omega_{even}} \right] \nonumber \\
		& \ & + {\cal O} (B^{2}({}^{0}\rho) \exp \left[ -2 \frac{\tau_{even}}{{}^{0}\Omega_{even}} \right]), \ \ \ \ \ \ 
\eea
where $B({}^{0}\rho)$ behaves for large $^{0}\rho$ like
\beq
\label{eq:Beigsconjecture2}
	B({}^{0}\rho) \sim \begin{rm}{const}\end{rm} \cdot \cosh{\left[ \frac{M^3}{3 {}^{0}C_{lim} {}^{0}\Omega_{even}} 
								\left( \frac{{}^{0}\rho}{M} \right)^{3} \right]}.
\eeq
Here the constant appearing in the equation for \hbox{$B({}^{0}\rho)$} is given by \hbox{$3 ^{0}C_{lim}$$^{0}\Omega_{even} / M^3 = \frac{81}{16}\sqrt{2}$}, hence note a misprint in the previously stated reference \cite{pc_Murchadha}.
The particular form of $B({}^{0}\rho)$ in (\ref{eq:Beigsconjecture2}) is motivated by the solution to the lapse equation, \hbox{$\triangle \alpha = R \alpha$}, on the limiting slice \hbox{$r = {}^{0}r_{C_{lim}} = \frac{3M}{2}$}.

We now want to look at the event horizon and first work out the late time behavior of the volume radius there to be denoted by $^{0}\rho^{+}_{C}$, the subscript $C$ marking its time dependence.
A good starting point for this task is Eq.~(\ref{eq:drhodC}), which using the chain rule, \hbox{$\frac{\partial \rho}{\partial \delta} = \frac{\partial \rho}{\partial C} \frac{dC}{d\delta}$}, can be written as
\bea
\label{eq:drhoplusdd}
           \frac{\partial {}^{0}\rho^{+}_{C}}{\partial \delta}
           & = & \frac{C}{{}^{0}\rho^{+\:2}_{C}} \left(\frac{\partial}{\partial C} H_{C}({}^{0}r_{+})\right) 
					   \frac{dC}{d\delta} \nonumber \\
	   & = & \frac{1}{{}^{0}\rho^{+\:2}_{C}} \Bigg(
						\frac{{}^{0}r^{2}_{+}}
						     {2({}^{0}r_{+} - \frac{3}{2}M)} 	
	        			      - \frac{1}{2} C K_{C}({}^{0}r_{+})
					        \Bigg) \frac{dC}{d\delta} \ \ \ \ \;
\eea
using (\ref{eq:dCdH}) evaluated at $^{0}r_{+}$.
To solve (in leading order) this ordinary differential equation (ODE), one can make use of \hbox{$\delta$-expansions} for $K_{C}(+)$, (\ref{eq:Kinfseries}) with (\ref{eq:KinfminusKplusseries}), $C$, (\ref{eq:Cseries}), and hence $\frac{dC}{d\delta}$, (\ref{eq:dCddproof}), derived in the Appendix.
It is important to keep in mind that $\delta$ at late times is decaying exponentially with time according to (\ref{eq:deltaevenRN}).
With the first term in the second set of large parentheses of (\ref{eq:drhoplusdd}) being of order ${\cal O}(1)$ only whereas \hbox{$K_{C}({}^{0}r_{+}) = {\cal O}(\frac{1}{\delta^{2}})$}, it turns out that 
\bea
\label{eq:rhoplusODE}
	^{0}\rho_{C}^{+\:2} \frac{d{}^{0}\rho^{+}_{C}}{d\delta} 
	& = & - \frac{1}{2} C K_{C}({}^{0}r_{+}) \frac{dC}{d\delta} + {\cal O}(\delta) \nonumber \\
        & = & - {}^{0}C_{lim} {}{}^{0}\Omega_{even} \:\frac{1}{\delta} + {\cal O}(1)  
\eea  
is found with \hbox{$^{0}C_{lim} = \frac{3}{4}\sqrt{3}M^{2}$}, (\ref{eq:defClim}).
We observe that in leading order
\beq
	^{0}\rho_{C}^{+}(\delta) \simeq \sqrt[3]{- 3 {}^{0}C_{lim} {}^{0}\Omega_{even} \ln{\left[\delta\right]}} 
		      \ \ \begin{rm}{as}\end{rm} \ \ 
      		      \delta \to 0
\eeq
and hence 
\beq
\label{eq:rhoplusmoves}
	^{0}\rho_{C}^{+}(\tau_{even}) \simeq \sqrt[3]{3{}^{0}C_{lim} \tau_{even}} 
                      \ \ \begin{rm}{as}\end{rm} \ \ 
      		      \tau_{even} \to \infty 
\eeq
hold.
So indeed the conjecture of \cite{Beig98} can be applied for the event horizon as the latter is found at large values of $^{0}\rho$ at late times.
This is a slice stretching effect as will be pointed out for similar coordinates in Subsecs.~\ref{subsec:LTAzeroeven} and E.

Now putting this result into the conjecture (\ref{eq:Beigsconjecture1}), we find in leading order 
\beq
	^{0}\alpha_{even}(\tau_{even},{}^{0}\rho_{C}^{+} (\tau_{even})) 
	\simeq {\cal O}(1),
\eeq 
so the lapse approaches a finite value there in the limit \hbox{$\tau_{even} \to \infty$}.
This clearly is in agreement with (\ref{eq:evenlapseREHRN}), which states that the lapse at the right-hand event horizon has to approach the finite value $\frac{3}{16}\sqrt{3}$ for maximal slicing of Schwarzschild with even boundary conditions.

\subsection{Isotropic grid coordinates}
\label{subsec:Gridcoordinates}

In order to compare with numerical output calculated for maximal slicing of puncture evolutions, we construct isotropic grid coordinates as in \cite{mypaper1}. To focus on the dynamical features of the \hbox{3-metric} rather than the static singularity at the puncture, we shall write
\beq
\label{eq:3mshift}
	^{(3)}ds^{2} = G\:dx^2 + r^{2}\:d\Omega^{2} 
                     = \Psi^{4} \left[ g\:dx^2 + \frac{r^{2}}{\Psi^{4}}\:d\Omega^{2}\right] ,
\eeq
introducing with small letters the rescaled \hbox{3-metric} $g_{ij}$ and its radial component $g$ by 
\beq
\label{eq:gijandg1}
	g_{ij}(C(\tau),x) = \frac{G_{ij}(C(\tau),x)}{\Psi^{4}(x)} 
\eeq
and
\beq
\label{eq:gijandg2}
        g(C(\tau),x) = \frac{G(C(\tau),x)}{\Psi^{4}(x)},
\eeq
respectively.
Here as pointed out in \cite{Anninos95} the conformal factor is determined on the initial slice from the \hbox{3-metric} written in isotropic coordinates. 
Hence $\Psi(x)$ is constant in time and given by (\ref{eq:conformalgamma}),
\beq
\label{eq:conformalpsi4}
	\Psi^{4} (x)  = \left[\left(1+\frac{M}{2x}\right)^{2} - \frac{qM^{2}}{4x^{2}} \right]^{2}.
\eeq
So initially
\beq
\label{eq:ginitials}
	G(C = \tau = 0,x) = \Psi^{4}(x)
        \ \ \begin{rm}{and}\end{rm} \ \ 
	g(C = \tau = 0,x) = 1
\eeq
and in addition
\beq
\label{eq:rinitials}
	r(C = \tau = 0,x) = x \Psi^{2}(x)
\eeq
has to hold since on the initial slice the radial grid coordinate $x$ is identical to the Schwarzschild isotropic coordinate $R$ defined in (\ref{eq:rofR}) in terms of the Schwarzschild radial coordinate $r$.

We want to emphasize that when odd boundary conditions are used in a numerical simulation, the isotropic lapse (\ref{eq:alphaisotropic}) together with the formulas (\ref{eq:ginitials}) and (\ref{eq:rinitials}) are valid at all times and the code has to reproduce the static Reissner-Nordstr\"om metric in isotropic coordinates.

For other boundary conditions, however, the components of the \hbox{4-metric} evolve with time.
Using that for vanishing shift the determinant of the \hbox{3-metric} has to be time-independent, by imposing the initial conditions (\ref{eq:ginitials}) and (\ref{eq:rinitials}), in \cite{mypaper1} we were able to find
\beq
\label{eq:g}
	g (C(\tau),x) = \frac{x^{4} \Psi^{8}(x)}{r^{4}(C(\tau),x)}.   
\eeq
It is from \hbox{$G = \Psi^{4} g = \frac{x^{4}\Psi^{12}}{r^{4}} = \gamma \left( \frac{\partial r}{\partial x} \right)^{2}$} then trivial to infer for fixed slice label $C$ the ODE
\beq
\label{eq:drdx}
	\frac{\partial r}{\partial x} 
            = \pm \frac{\sqrt{p_{C}(r)}}{r^{4}} x^{2} \Psi^{6}(x)
\eeq
relating the Schwarzschild radial coordinate $r$ to the radial grid coordinate $x$.
This equation can be integrated using the throat as lower integration limit by
\beq
\label{eq:ximplicitofr}
	\int\limits_{r_{C}}^{r} \frac{y^{4}\:dy}{\sqrt{p_{C}(y)}} 
  = \pm \int\limits_{x_{C}}^{x} y^{2} \Psi^{6}(y) \:dy,
\eeq
where $x_{C}$ denotes the location of the throat on the grid as a function of time and the ``+'' or ``-'' sign applies for the right-hand or left-hand side of the throat, respectively. 
One may readily check that for \hbox{$C = 0$} the grid coordinates coincide with the Schwarzschild isotropic coordinates where the throat is found at $x_{+}$.
For later times, though, $x_{C}$ has to be found from (\ref{eq:ximplicitofr}) by demanding that the coordinate transformation \hbox{$r = r(C(\tau),x)$} is consistent with the requirement of a vanishing shift.
The latter can be written as
\beq
	B = \left( \beta + \gamma \frac{\partial r}{\partial \tau} \right) \frac{\partial r}{\partial x} \equiv 0.
\eeq
Remembering the formula (\ref{eq:betafinal}) for $\beta$ and writing the lapse as \hbox{$\alpha = \frac{\sqrt{p_C(r)}}{r^2} \frac{\partial t}{\partial C} \frac{dC}{d\tau}$}, we observe that 
\beq
\label{eq:drdtau1}
	\frac{\partial r}{\partial \tau} = - \alpha \frac{C}{r^{2}} 
                                         = - \frac{\sqrt{p_C(r)} C}{r^4} \frac{\partial t}{\partial C} \frac{dC}{d\tau} 
\eeq
has to hold.
By differentiating (\ref{eq:ximplicitofr}) with respect to $C$, which making use of (\ref{eq:dCdH}), (\ref{eq:dCdHmHisI}) and (\ref{eq:drCdC}) yields
\bea		
\label{eq:getdrdC}
	  \frac{\partial r}{\partial C} \frac{r^4}{\sqrt{p_C(r)}}
      	      & + & C \Big( \frac{r^{2}}{2(r - \frac{3}{2}M + \frac{qM^{2}}{2r})\sqrt{p_C(r)}} \Big. \nonumber \\ 
	      & - & \Big. \frac{1}{2} K_{C}(r) \Big) 
	        = \mp x_{C}^{2} \Psi^{6}(x_{C}) \frac{dx_{C}}{dC},
\eea
we obtain 
\bea
\label{eq:drdtau2}
  	 \frac{\partial r}{\partial \tau} 
	      & = & - \frac{\sqrt{p_C(r)} C}{r^4} \Big(
		\frac{r^{2}}{2(r - \frac{3}{2}M + \frac{qM^{2}}{2r})\sqrt{p_C(r)}} \Big. \nonumber \\
              & - & \Big. \frac{1}{2} K_{C}(r)
		\pm \frac{1}{C} x_{C}^{2} \Psi^{6}(x_{C}) \frac{dx_{C}}{dC} \Big) \frac{dC}{d\tau}
\eea
as a further equation for $\frac{\partial r}{\partial \tau}$.
We want to point out that since in (\ref{eq:drdtau1}) the lapse enters, the behavior of the throat, to be determined by a comparison of (\ref{eq:drdtau1}) and (\ref{eq:drdtau2}), depends on the boundary conditions imposed on the lapse.  
In Subsecs.~\ref{subsec:LTAzeroeven} and \ref{subsec:LTAzerozgp} we show this concept for even and ``zgp'' boundary conditions explicitly when analyzing slice stretching effects in the limit of late times.
A study of slice stretching in the context of more general boundary conditions is the subject of \cite{mypaper3}.

\subsection{Slice stretching effects for even boundary conditions}
\label{subsec:LTAzeroeven}

\subsubsection{Throat $x_C$}
Comparing (\ref{eq:drdtau1}) and (\ref{eq:drdtau2}), for even boundary conditions with the lapse given by (\ref{eq:alphaevenfinal}) or the partial derivative of the height function with respect to $C$ by (\ref{eq:dCdH}), we observe that 
\beq
\label{eq:eventhroatfix}
	\frac{dx_C}{dC} = 0
\eeq
holds and the throat hence remains at \hbox{$x_{C = 0} = x_+$}.
Remembering that $\delta$ for even boundary conditions has been found to decay exponentially with $\tau_{even}$ on the fundamental time scale $\Omega$, (\ref{eq:deltaevenRN}), we observe that the \hbox{3-metric} ``freezes'' quickly at the throat and with (\ref{eq:g}) for the rescaled radial metric component
\beq
	0 < \lim_{\tau_{even} \to \infty} g(\tau_{even},x_+) 
          = \frac{x_+^4 \Psi^8(x_+)}{r^4_{C_{lim}}} < \infty,
\eeq
and for the angular part
\beq
	0 < \lim_{\tau_{even} \to \infty} 
	    \frac{r^{2}(\tau_{even},x_{+})}{\Psi^{4}(x_{+})} 
          = \frac{r_{C_{lim}}^2}{\Psi^{4}(x_+)} < \infty
\eeq
are found.
In particular, we want to point out that for the Schwarzschild spacetime $g$ at the throat only grows up to the finite value \hbox{$(\frac{4}{3})^4 \approx 3.1605$}.

\subsubsection{Event horizon at $x_C^+$ and $x_C^-$}
When discussing the location of the event horizon, note that with the throat remaining at $x_+$ for all times, in (\ref{eq:ximplicitofr}) the isometry (\ref{eq:isometry}) is present which maps the left-hand side of the throat to the right-hand side and vice versa.

In order to derive an ODE for the location of the right-hand event horizon, we start by differentiating (\ref{eq:ximplicitofr}) with the integration limit given by the right-hand event horizon, i.e.\ the pair \hbox{$\{r_{+}, x_{C}^{+}\}$}.
By a calculation similar to the one leading to (\ref{eq:drhodC}) or (\ref{eq:getdrdC}), it follows that 
\bea
\label{eq:evendxcpdC1}
      	  \frac{r_{+}^{2}}{2(r_{+} - \frac{3}{2}M + \frac{qM^{2}}{2r_{+}})} 
	      & - & \frac{1}{2} C K_{C}(r_{+}) \nonumber \\
	      & = & x_{C}^{+\:2} \Psi^{6}(x_{C}^{+}) \frac{dx_{C}^{+}}{dC}. \ \ \ \ \ 
\eea
Using the chain rule, \hbox{$\frac{dx_{C}}{dC} = \frac{dx_{C}}{d\delta} \frac{d\delta}{dC}$}, and \hbox{$\delta$-expansions} for $K_{C}(r_+)$, (\ref{eq:Kinfseries}) with (\ref{eq:KinfminusKplusseries}), $C$, (\ref{eq:Cseries}), and hence $\frac{dC}{d\delta}$, (\ref{eq:dCddproof}), derived in the Appendix, we find in leading order of $\delta$ the ODE
\beq
\label{eq:evendxcpdC2}
	- C_{lim} \Omega \:\frac{1}{\delta} + {\cal O}(1)
        = x_{C}^{+\:2} \Psi^{6}(x_{C}^{+}) \frac{dx_{C}^{+}}{d\delta}.
\eeq 
Note the similarity to the ODE (\ref{eq:rhoplusODE}) obtained for volume radius coordinates.
There we found that the event horizon moves to the right toward infinity, and in this limit the conformal factor approaches one. 
Presupposing this result and setting the conformal factor appearing in (\ref{eq:evendxcpdC2}) to unity,
\beq
	- C_{lim} \Omega \:\frac{1}{\delta} 
        \simeq x_{C}^{+\:2} \frac{dx_{C}^{+}}{d\delta},
\eeq 
we infer that the right-hand event horizon expanded in $\delta$ as
\beq
\label{eq:evenxcplusofdelta}
	x_{C}^{+}(\delta) \simeq \sqrt[3]{- 3 C_{lim} \Omega \ln{\left[\delta\right]}} 
	              \ \ \begin{rm}{as}\end{rm} \ \ 
                      \delta \to 0 
\eeq
moves at late times to infinite values of $x$ like
\beq
\label{eq:evenxcplust}
	  x_{C}^{+}(\tau_{even}) \simeq \sqrt[3]{3 C_{lim} \tau_{even}} 
        	      \ \ \begin{rm}{as}\end{rm} \ \ 
                      \tau_{even} \to \infty . 
\eeq
Note that the right-hand event horizon is found on the grid at a location that in leading order is proportional to the third root of the time.

Furthermore, making use of the isometry, the left-hand event horizon is found to move toward the puncture according to
\beq
\label{eq:evenxcminust}
	  x_{C}^{-}(\tau_{even}) \simeq \frac{(1-q)M^2}{4 \sqrt[3]{3 C_{lim} \tau_{even}}} 
	           	         \ \ \begin{rm}{as}\end{rm} \ \ 
        	                 \tau_{even} \to \infty .  
\eeq
For $g$ at both the left-hand and right-hand event horizon in leading order the statement 
\beq
    g(\tau_{even},x_{C}^{\pm}(\tau_{even})) \simeq \frac{\left( 3 C_{lim} \tau_{even}\right)^{\frac{4}{3}}}
                                    {r_{+}^{4}}
        	      \ \ \begin{rm}{as}\end{rm} \ \ 
                      \tau_{even} \to \infty 
\eeq
can be made.
Hence at the event horizon the rescaled radial metric component is proportional to $\tau_{even}^{4/3}$ and grows without bounds.  
For the angular part we note that whereas on the right-hand side the expression
\beq
	\lim_{\tau_{even} \to \infty} \frac{r^{2}(\tau_{even},x^{+}_{C}(\tau_{even}))}{\Psi^{4}(x^{+}_{C}(\tau_{even}))} 
        = r_{+}^{2} 
\eeq
remains finite, on the left-hand side the corresponding term,
\beq
	\lim_{\tau_{even} \to \infty} \frac{r^{2}(\tau_{even},x^{-}_{C}(\tau_{even}))}{\Psi^{4}(x^{-}_{C}(\tau_{even}))} = 0, 
\eeq
approaches zero as the conformal factor grows without bound.

\subsection{Slice stretching effects for zgp boundary conditions}
\label{subsec:LTAzerozgp}
\begin{figure}[ht]
	\noindent
	\epsfxsize=85mm \epsfysize=185mm \epsfbox{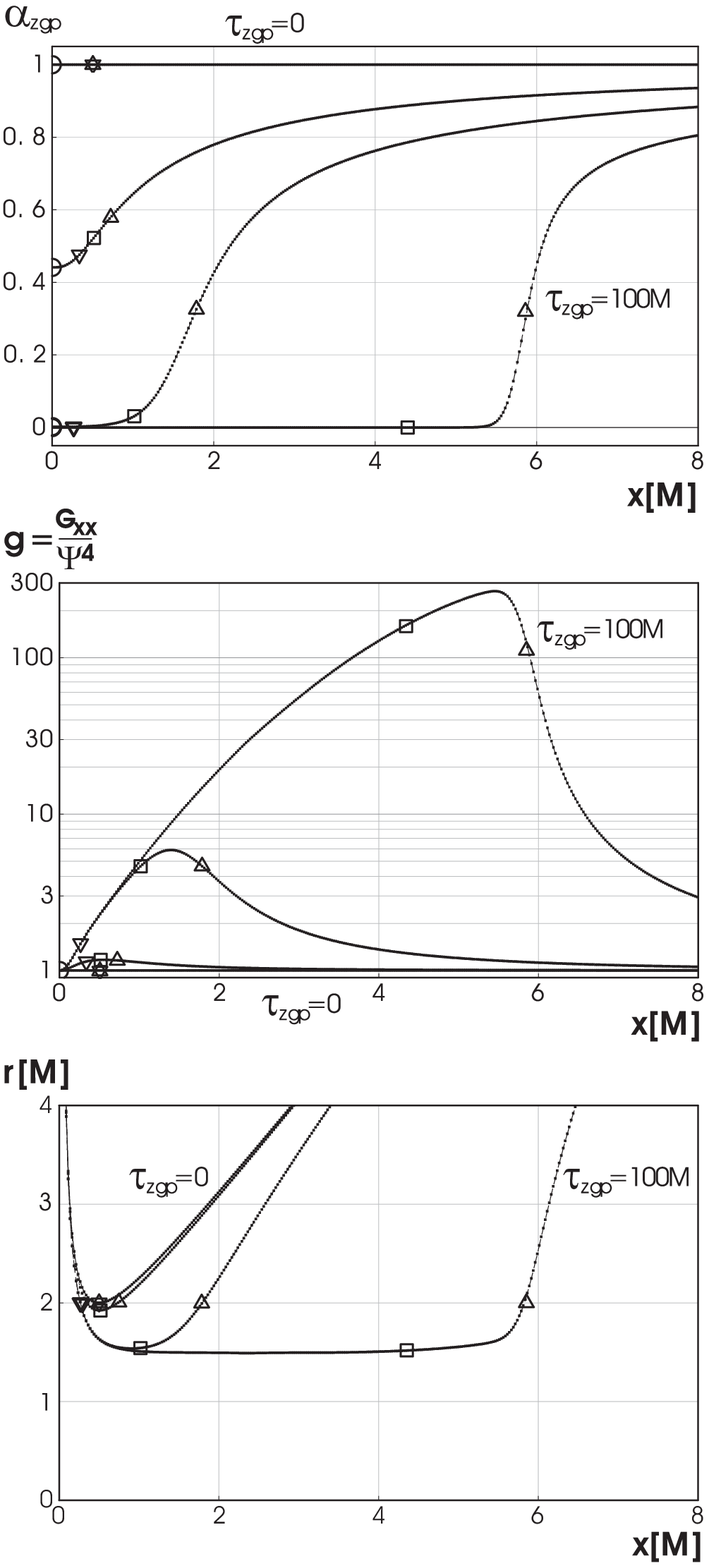}
	\caption{For ``zgp'' boundary conditions numerical results at times \hbox{$\tau_{zgp} = \left\{ 0, 1M, 10M, 100M \right\}$} are shown to demonstrate the ``collapse of the lapse'' and ``slice wrapping and sucking'' taking place as the limiting slice \hbox{$r = \frac{3M}{2}$} is approached. The ``symptoms'' of slice stretching are the development of a rapidly growing peak in the radial metric function (note the logarithmic scale!) and the outward-drifting of coordinate locations. To study the latter, one should observe the location of the throat, the peak in the radial component of the rescaled \hbox{3-metric}, and the right-hand event horizon denoted by boxes, stars, and upward pointing triangles, respectively.}
\label{fig:Metric}	
\end{figure}

Making use of the puncture data line element, (\ref{eq:3mshift}) together with (\ref{eq:g}), and the coordinate transformation relating the radial gauge to isotropic grid coordinates, (\ref{eq:ximplicitofr}), in \cite{mypaper1} we were able to verify the analytically derived 4-metric with numerical output for maximal slicing of a Schwarzschild black hole with ``zgp'' boundary conditions and vanishing shift.
In Fig.~\ref{fig:Metric} we show $\alpha_{zgp}$, $g$ (plotted on a logarithmic scale) and $r$ obtained numerically at times \hbox{$\tau_{zgp}=\{0,1M,10M,100M\}$}.
The reader should observe that with the throat and the left-hand and right-hand event horizon acting as markers on the profiles of the \hbox{3-metric} (the ``open boxes'' and ``triangles'' in Fig.~\ref{fig:Metric}), by determining their locations on the grid as a function of time and evaluating the \hbox{3-metric} at those locations, one can describe in the late time limit the effects of slice stretching.

\subsubsection{Throat $x_C$}
For ``zgp'' boundary conditions, the ODE which determines the location of the throat is most easily derived by comparing the partial derivative with respect to $C$ of the ``zgp'' height function, (\ref{eq:zgpdtdC}), appearing in (\ref{eq:drdtau1}) with the term in the brackets of (\ref{eq:drdtau2}).
This yields 
\beq
\label{eq:throatODEorigin}
	- \frac{1}{2} C ( K_{C}(\infty) + \frac{2}{M} )
         = x_{C}^{2} \Psi^{6}(x_{C}) \frac{dx_{C}}{dC}.   
\eeq  
Therefore, for $x_{C}(\delta)$ based on \hbox{$\delta$-expansions} for $K_{C}(\infty)$, (\ref{eq:Kinfseries}), $C$, (\ref{eq:Cseries}), and hence $\frac{dC}{d\delta}$, (\ref{eq:dCddproof}), in leading order of $\delta$ the ODE
\beq
\label{eq:throatODE}
        - C_{lim} \Omega \:\frac{1}{\delta} + {\cal O}(1)
        = x_{C}^{2} \Psi^{6}(x_{C}) \frac{dx_{C}}{d\delta}   
\eeq  
is obtained.
Note the similarity of this ODE for the outward-movement of the throat in isotropic grid coordinates at late times to the ODE (\ref{eq:rhoplusODE}) describing the location of the event horizon in volume radius coordinates and to (\ref{eq:evendxcpdC2}) for the right-hand event horizon in the context of even boundary conditions.
Neglecting again the conformal factor in order to infer the behavior of $x_{C}(\delta)$ in the limit of late times, by solving
\beq
\label{eq:modelzeroshift}
	- C_{lim} \Omega \:\frac{1}{\delta} \simeq x_{C}^{2} \frac{dx_{C}}{d\delta}
\eeq
we find 
\beq
\label{eq:xcofdelta}
	x_{C}(\delta) \simeq \sqrt[3]{- 3 C_{lim} \Omega \ln{\left[\delta\right]}}
	              \ \ \begin{rm}{as}\end{rm} \ \ 
                      \delta \to 0 .
\eeq
Since $\delta$ is decaying exponentially with $\tau_{zgp}$ according to (\ref{eq:deltazgpRN}) on the time scale $\Omega_{zgp} = 2 \Omega$, (\ref{eq:OmegazgpRNis}), the location of the throat on the grid at late times is hence dominated by 
\beq
\label{eq:xct}
	  x_{C}(\tau_{zgp}) \simeq \sqrt[3]{\frac{3}{2} C_{lim} \tau_{zgp}} 
                      \ \ \begin{rm}{as}\end{rm} \ \ 
      		      \tau_{zgp} \to \infty.
\eeq
Of course, this simple solution is not expected to give a good approximation at early times since both the expansion in $\delta$ and neglecting the term $\Psi^{6}(x_{C})$ are not valid in this regime.
But at late times the numerically computed outward movement of the throat agrees well with the power law (\ref{eq:xct}) obtained for Schwarzschild and ``zgp'' boundary conditions, c.f.\ Fig.~\ref{fig:locblow}.

Since we are interested in the late time behavior of the \hbox{3-metric}, bearing in mind the definition of $\delta$, (\ref{eq:defdeltaRN}), we can infer from (\ref{eq:g}) that the rescaled radial component of the metric at the throat blows up as $\tau_{zgp}^{4/3}$, i.e.\
\beq
\label{eq:gxct}
	g(\tau_{zgp},x_{C}(\tau_{zgp})) \simeq  \frac{\left( \frac{3}{2} C_{lim} \tau_{zgp} \right)^{\frac{4}{3}}}
                                    {r_{C_{lim}}^{4}}
        	      \ \ \begin{rm}{as}\end{rm} \ \ 
                      \tau_{zgp} \to \infty.
\eeq
The angular part, though, remains finite,
\beq
	0 < \lim_{\tau_{zgp} \to \infty} \frac{r^{2}(\tau_{zgp},x_{C}(\tau_{zgp}))}{\Psi^{4}(x_{C}(\tau_{zgp}))} 
          = r_{C_{lim}}^{2} < \infty.
\eeq

\subsubsection{Right-hand event horizon $x_C^+$}
Now differentiating (\ref{eq:ximplicitofr}) with the integration limit being the right-hand event horizon, i.e.\ the pair \hbox{$\{r_{+}, x_{C}^{+}\}$}, with respect to $C$, it follows that
\bea
\label{eq:dxcpdC}
      	      & \ & \frac{r_{+}^{2}}{2(r_{+} - \frac{3}{2}M + \frac{qM^{2}}{2r_{+}})} 
	      	    - \frac{1}{2} C K_{C}(r_{+}) \nonumber \\
	      & = & x_{C}^{+\:2} \Psi^{6}(x_{C}^{+}) \frac{dx_{C}^{+}}{dC} - x_{C}^{2} \Psi^{6}(x_{C}) \frac{dx_{C}}{dC}. \ \ \ \ \ 
\eea
Adding the equations (\ref{eq:throatODEorigin}) and (\ref{eq:dxcpdC}) and using \hbox{$K_{C}(\infty) - K_{C}(r_{+}) = \eta + {\cal O} (\delta^{2})$}, (\ref{eq:KinfminusKplusseries}), we find in leading order of $\delta$ the ODE
\beq
	- 2 C_{lim} \Omega \:\frac{1}{\delta} + {\cal O}(1)
        = x_{C}^{+\:2} \Psi^{6}(x_{C}^{+}) \frac{dx_{C}^{+}}{d\delta}.
\eeq 
Note that this ODE for the location of the right-hand event horizon at late times only differs by a factor of two from (\ref{eq:throatODE}) obtained for the throat.
As for the throat we set the conformal factor to one and infer from
\beq
	- 2 C_{lim} \Omega \:\frac{1}{\delta} 
        \simeq x_{C}^{+\:2} \frac{dx_{C}^{+}}{d\delta}
\eeq 
that the right-hand event horizon expanded in $\delta$ as
\beq
\label{eq:xcplusofdelta}
	x_{C}^{+}(\delta) \simeq \sqrt[3]{- 6 C_{lim} \Omega \ln{\left[\delta\right]}} 
	              \ \ \begin{rm}{as}\end{rm} \ \ 
                      \delta \to 0 
\eeq
moves at late times to infinite values of $x$ like
\beq
\label{eq:xcplust}
	  x_{C}^{+}(\tau_{zgp}) \simeq \sqrt[3]{3 C_{lim} \tau_{zgp}} 
                      \ \ \begin{rm}{as}\end{rm} \ \ 
      		      \tau_{zgp} \to \infty, 
\eeq
i.e.\ with a location on the grid again proportional to the third root of the time. 
However looking at the ``collapse speed'', i.e.\ the rate at which the numerical grid falls into the black hole, one can see from
\beq
   \lim_{\tau_{zgp} \to \infty} \frac{\frac{dx_{C}^{+}(\tau_{zgp})}{d\tau_{zgp}}}{\frac{dx_{C}(\tau_{zgp})}{d\tau_{zgp}}}
   \simeq \sqrt[3]{2} 
\eeq 
that the right-hand event horizon moves at late times by a factor of $\sqrt[3]{2}\approx 1.2599$ faster than the throat. 
For the 3-metric,
\beq
\label{eq:gxcplust}
    g(\tau_{zgp},x_{C}^{+}(\tau_{zgp})) \simeq \frac{\left( 3 C_{lim} \tau_{zgp}\right)^{\frac{4}{3}}}
                                    {r_{+}^{4}}
        	      \ \ \begin{rm}{as}\end{rm} \ \ 
                      \tau_{zgp} \to \infty 
\eeq
and
\beq
	0 < \lim_{\tau_{zgp} \to \infty} \frac{r^{2}(\tau_{zgp},x^{+}_{C}(\tau_{zgp}))}{\Psi^{4}(x^{+}_{C}(\tau_{zgp}))} 
          = r_{+}^{2} < \infty 
\eeq
can be found.
So the unbounded growth of $g$ is again proportional to $\tau_{zgp}^{4/3}$ and the radial part remains finite. 
Calculating
\beq
	\lim_{\tau_{zgp} \to \infty} \frac{\frac{dg(\tau_{zgp},x_{C}^{+}(\tau_{zgp}))}{d\tau_{zgp}}}{\frac{dg(\tau_{zgp},x_{C}(\tau_{zgp}))}{d\tau_{zgp}}}
   	\simeq 2^{\frac{4}{3}} \left( \frac{r_{C_{lim}}}{r_{+}} \right)^{4}
\eeq
we claim that for Schwarzschild the blowup of $g$ should occur for the right-hand event horizon by a factor \hbox{$2^{\frac{4}{3}} \left( \frac{3}{4} \right)^{4} \approx 0.7973$} slower than for the throat.  

\subsubsection{Left-hand event horizon $x_C^-$}
For the left-hand event horizon one has to take care of the minus sign on the right-hand side of (\ref{eq:ximplicitofr}).
As for the right-hand event horizon, with 
\bea
\label{eq:dxcmdC}
	       & \ & \frac{r_{+}^{2}}{2(r_{+} - \frac{3}{2}M + \frac{qM^{2}}{2r_{+}})} 
	      	     - \frac{1}{2} C K_{C}(r_{+}) \nonumber \\
               & = & - x_{C}^{-\:2} \Psi^{6}(x_{C}^{-}) 
                  	\frac{dx_{C}^{-}}{dC} + x_{C}^{2} \Psi^{6}(x_{C}) \frac{dx_{C}}{dC} \ \ \ \ 
\eea
subtracted from (\ref{eq:throatODEorigin}) we obtain 
\bea
\label{eq:dxcmdCsimple}
	  - \left( \frac{r_{+}^{2}}
                      {2(r_{+} - \frac{3}{2}M + \frac{qM^{2}}{2r_{+}})} \right.
             & + & \left. \frac{1}{2} C_{lim} (\eta - \frac{2}{M}) \right) 
		\frac{M -\mu}{\sqrt{-\sigma_{+}\sigma_{-}}} \delta  \nonumber \\
             + {\cal O}(\delta^2) 
             & = & x_{C}^{-\:2} \Psi^{6}(x_{C}^{-}) \frac{dx_{C}^{-}}{d\delta}. 
\eea 
As the prefactor of $\delta$ on the left-hand side of (\ref{eq:dxcmdCsimple}) is positive, we observe that the value of $x_{C}^{-}$ is decreasing as $\delta$ is decaying to zero.
For this reason we do not set the conformal factor to unity here but observe that in the limit of late times the left-hand event horizon will be found at a finite location on the grid to the left of its initial location, i.e.\ 
\beq
	0 < \lim_{\tau_{zgp} \to \infty} x_{C}^{-}(\tau_{zgp}) = x_{C_{lim}}^{-}  < x_+.
\eeq
It then follows that the \hbox{3-metric} ``freezes'' as 
\beq
\label{eq:gxcm}
          0 < \lim_{\tau_{zgp} \to \infty} g(\tau_{zgp},x_{C}^{-}(\tau_{zgp})) 
          = \frac{x_{C_{lim}}^{-\:4} \Psi^{8}(x_{C_{lim}}^{-})}{r_{+}^{4}} < \infty ,
\eeq
and
\beq
\label{eq:rxcm}
          0 < \lim_{\tau_{zgp} \to \infty} \frac{r^{2}(\tau_{zgp},x_{C}^{-}(\tau_{zgp}))}
                                      {\Psi^{4}(x_{C}^{-}(\tau_{zgp}))} 
          = \frac{r_{+}^{2}}{\Psi^{4}(x_{C_{lim}}^{-})} < \infty 
\eeq
holds. 

\subsubsection{Peak in the radial metric component at $x_C^p$}
\label{subsubsec:peak}
Finally we want to discuss the peak occurring in the radial component of the rescaled \hbox{3-metric} at location $x_C^p$.
Inferring from Fig.~\ref{fig:Metric} that $x_C^p$ is found to lie in between the location of the throat and the right-hand event horizon, 
\beq
	x_C \leq x_C^p \leq x_C^+,
\eeq
we claim that upper and lower bounds for the location of the peak are given by (\ref{eq:xct}) and (\ref{eq:xcplust}).
Furthermore, the corresponding value of the Schwarzschild radius is then bounded from below by $r_{C_{lim}}$ and from above by $r_+$.

From (\ref{eq:g}) we can therefore infer the following bounds for the peak of $g$, 
\beq
	\frac{\left( \frac{3}{2} C_{lim} \tau_{zgp}\right)^{\frac{4}{3}}}
                                    {r_{+}^{4}}
	\lesssim \ g(\tau_{zgp},x_C^p(\tau_{zgp})) \ \lesssim
        \frac{\left( 3 C_{lim} \tau_{zgp}\right)^{\frac{4}{3}}}
                                    {r_{C_{lim}}^{4}},
\eeq
which - setting for simplicity the conformal factor to 1 - have to hold in leading order at late times.

\subsection{Numerical results}
\label{subsec:SliceStretchingNum}

\begin{figure}[ht]
	\noindent
	\epsfxsize=85mm \epsfysize=120mm \epsfbox{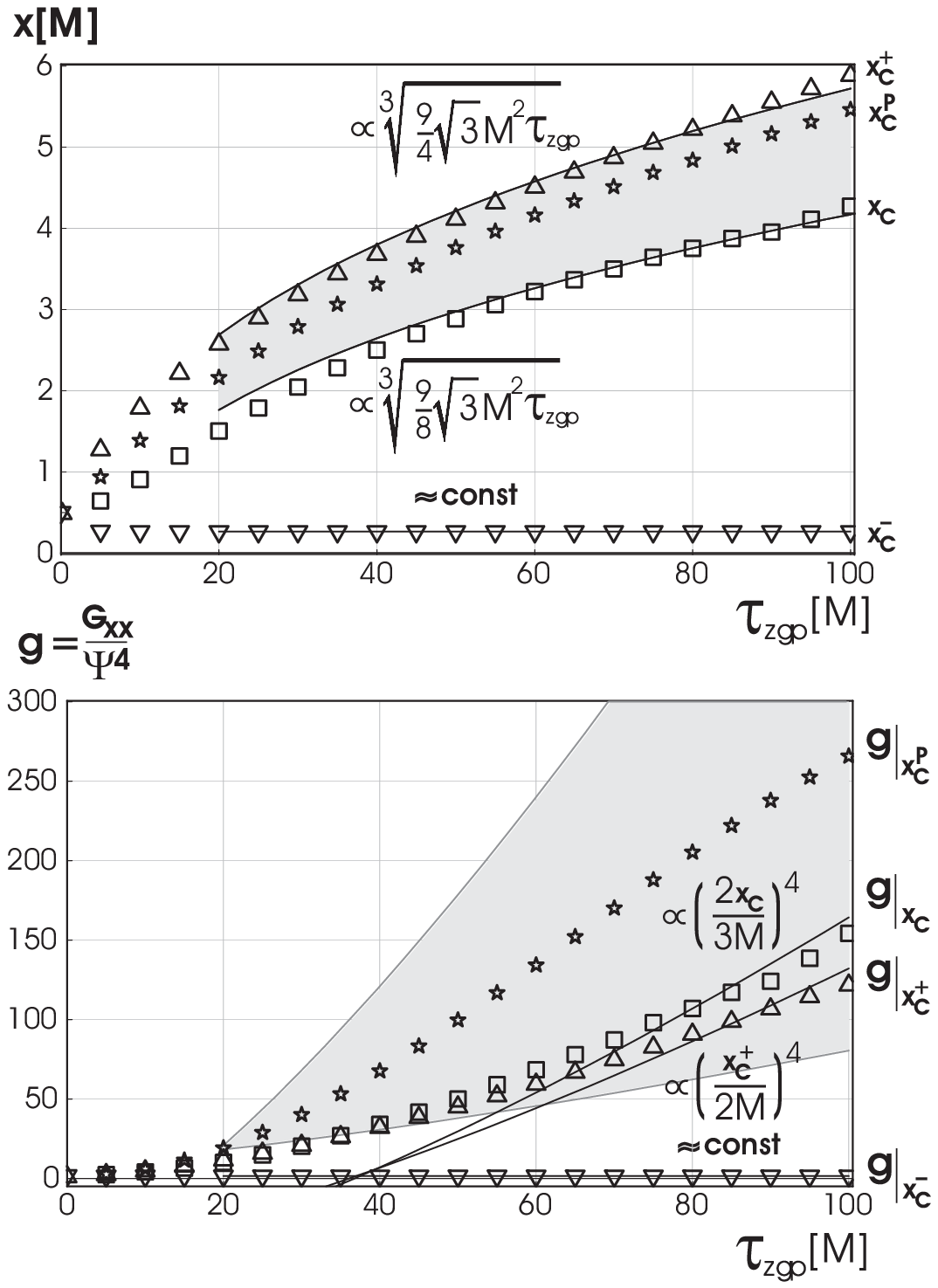}
	\caption{\textit{Top:} Slice sucking is studied numerically by plotting as a function of time the location on the grid of the throat (shown as boxes) and the right-hand event horizon (denoted by upward pointing triangles). In between the latter two the peak in the radial metric component (plotted as stars) is found. The analytically derived late time behavior (fitted at \hbox{$\tau_{zgp} = 80M$} and predicting a growth proportional to $\tau_{zgp}^{1/3}$ with known prefactors) agrees very well with numerical results.  \textit{Bottom:} Similarly slice wrapping can be investigated by evaluating the radial part of the 3-metric at those locations. At late times again agreement with the analytically derived blow-up proportional to $\tau_{zgp}^{4/3}$ is found. The gray area indicates the bounds obtained for the peak in $g$.}
\label{fig:locblow}	
\end{figure}

The late time statements about slice stretching for vanishing shift have been compared for ``zgp'' boundary conditions and vanishing charge to numerical simulations, see Sec.~\ref{subsec:LTACactus} for comments on the numerical details.
We summarize the result in Fig.~\ref{fig:locblow}. 
Note that this data helps explain e.g.\ Fig.~3.13 for the outward movement of the event horizon and Fig.~3.16 for the growing peak in the radial metric function $g$ in \cite{Camarda97}.

To determine the numerical data points of Fig.~\ref{fig:locblow}, the locations of the throat and the left-hand and right-hand event horizon on a given slice had to be determined. 
Whereas an event horizon can be located by searching for the isosurfaces \hbox{$r = \mbox{}^{0}r_{+} = 2M$}, searching for the throat as the minimum of $r$ is difficult at late times as the minimum is very shallow, c.f.\ Fig.~\ref{fig:Metric}.
Instead the throat has been located via its value of the lapse function that is given at late times by (\ref{eq:zgplapsethroatRN}).
Finding the event horizon by the corresponding values of the lapse yields locations that are very similar to the ones obtained with the isosurface based method.

Note in Fig.~\ref{fig:locblow} that for the left-hand event horizon both the location and the value of $g$ are essentially constant in agreement with analytical predictions. 
In addition, one can infer for the throat and the right-hand event horizon that the late time behavior for the locations is very well reproduced by (\ref{eq:xct}) and (\ref{eq:xcplust}).
The corresponding curves have been fitted to the numerical value at \hbox{$\tau_{zgp} = 80M$} since the LTA only reproduces the behavior at late times but is not capable of describing the behavior at early times.
Fitting at this value of $\tau_{zgp}$ is a compromise between applying the LTA at reasonably late times and avoiding large numerical errors at later times.

For the blowup of $g$ the accelerating growth taking place at the throat and the right-hand even horizon is reproduced qualitatively by (\ref{eq:gxct}) and (\ref{eq:gxcplust}). 
However, the reader should bear in mind several factors which contribute to deviations.
On the one hand several simplifications, valid only at late times, have been used when deriving the late time statements.
On the other hand the numerical errors in determining values at only approximately known positions by interpolation in a sharply peaked metric profile are not negligible. 
Furthermore, the numerical outer boundary is an approximation to a radiative boundary condition, and its influence is felt by the interior solution at later times, producing an error in the height of the metric peak.

Finally, Fig.~\ref{fig:locblow} also shows the bounds for the peak in the radial metric component derived in Sec.~\ref{subsubsec:peak}.
These bounds are not very sharp, c.f.\ the gray area in Fig.~\ref{fig:locblow}.
But note that the peak in the radial metric component has to grow proportional to $\tau_{zgp}^{4/3}$. An exponential growth of the \hbox{3-metric}, which in numerical relativity was sometimes expected for the slice stretching problem,
can be ruled out.

\section{Conclusion and Outlook}
\label{sec:conclusion}
In this paper we have considered the late time limit of maximal slicing of the extended Reissner-Nordstr\"om spacetime for both even and numerically motivated ``zgp'' boundary conditions.
In this limit, the leading order terms of the lapse function and the \hbox{3-metric} at the puncture, the left-hand event horizon, the throat, and the right-hand event horizon together with the corresponding locations on the grid as a function of time have been obtained for vanishing shift.

Taken together, the late time analysis for the lapse and statements for the grid locations provide an analytic understanding of the numerically observed lapse profile for ``zgp'' boundary conditions with the ``collapse of the lapse'' and the ``outward moving shoulder''.
The lapse has been shown to collapse at late times at the puncture, the left-hand event horizon and the throat, with the latter moving outward. The value of the lapse at the right-hand event horizon, however, moving outward slightly faster than the throat, approaches the finite value \hbox{$0 < \frac{C_{lim}}{r_{+}^{2}} \leq \frac{{}^{0}C_{lim}}{{}^{0}r_{+}^{2}} = \frac{3}{16}\sqrt{3} \approx 0.3248$}. 

Furthermore, we have discussed slice stretching effects, which have unpleasant consequences in numerical simulations.
The associated effects, such as outward-drifting of the throat and the right-hand event horizon (slice sucking) and unbounded growth of components of the 3-metric (slice wrapping), have been described analytically.
In particular, we were able to rule out an exponential growth of the latter.

For the ``zgp'' Schwarzschild case, the analytic statements for the outward moving shoulder in the lapse profile and the development of a rapidly growing peak in the rescaled radial metric function have been verified numerically.

Note that for constant mean curvature slicing of the Schwarzschild spacetime the time scale of the collapse of the lapse has been determined as a function of $K$ in \cite{Malec03}. 
It would be interesting to extend our studies in this direction.

For more accurate and longer lasting simulations with singularity avoiding slicing, promising strategies are either to use singularity excision or to use a non-trivial shift, which as the slice stretching develops reacts by pulling out grid points from the inner region. 
This has to be done based on appropriate conditions imposed on the 3-metric.
Candidates like the minimal distortion shift have been studied numerically as early as e.g.\ \cite{Bernstein89}. 
The Gamma-freezing shift conditions discussed in \cite{Alcubierre00c,Alcubierre01,Alcubierre02a} have allowed evolution times of $1000M$ and more, approximately freezing slice stretching. 
It is therefore of significant interest to provide an analytic understanding of the late time behavior of these shift conditions as is now available for maximal slicing.
Some analytical work has been already done in \cite{mythesis}, and we plan to examine shift conditions in more detail in the future.

\bigskip
\acknowledgments
It is a pleasure to thank M.~Alcubierre, R.~Beig, N.~\'{O} Murchadha, D.~Pollney, and E.~Seidel for discussions. 
This work was supported by NSF grant PHY-02-18750.
We also acknowledge the support of the Center for Gravitational Wave Physics funded by the National Science Foundation under Cooperative Agreement PHY-01-14375.

\appendix
\section*{\ \ \ \ \ \ \ \ \ \ \ \ \ \ Appendix: \newline Proof of the late time analysis}

\label{appendix:proof}
\subsection{Ansatz}
For our LTA, in contrast to \cite{Beig98} we will not rescale quantities like $C, \tau$ or $\delta$ by their units of $M$ here, since quite often the involved powers of $M$ give insight into the underlying physics and provide a first check of the formulas. 
So throughout this proof $M$ will be understood as a positive constant not necessarily equal to one, \hbox{$M>0$}. 
Furthermore, we will restrict the charge parameter \hbox{$q = \frac{Q^{2}}{M^{2}}$}, the squared ratio between charge $Q$ and mass $M$, to \hbox{$0 \leq q < 1$} to discuss only the non-extremal Reissner-Nordstr\"om spacetime. 
   
The ansatz for the LTA is based on an expansion in \hbox{$\delta = r_{C} - r_{C_{lim}}$}, (\ref{eq:defdeltaRN}), i.e.\ the difference between the radial coordinate at the throat $r_{C}$ on a slice labeled with $C$ and its limiting value $r_{C_{lim}}$ corresponding to $C_{lim}$.
As $\delta$ for the non-extremal Reissner-Nordstr\"om metric decreases strictly monotonically for \hbox{$C \in \left[ 0,C_{lim} \right]$}, the range of $\delta$ is \hbox{$\delta \in [0,\sigma_{+}]$}, and the late time limit is obtained by considering the limit \hbox{$\delta \to 0$}.
Here for convenience \hbox{$\sigma_{\pm} \:^{>}_{<} 0$} as in (\ref{eq:SIGMA}) has been introduced. 
Note that $\sigma_{+}$ as a function of $q$ decreases from $\frac{M}{2}$ for the Schwarzschild case to zero for the extremal Reissner-Nordstr\"om case, where the LTA obviously breaks down.

We want to point out that, for the non-extremal Reissner-Nordstr\"om metric, $\delta$ defined by (\ref{eq:defdeltaRN}) yields a valid ansatz for the discussion of the late time behavior of spacelike maximal slices extending on both sides of the throat to infinity (referred to as ``horizon-horizon'' slices in \cite{Gentle2001}) unless one starts with the initial slice being the limiting slice \hbox{$r = r_{C_{lim}}$}.

The task of finding the \hbox{$\delta$-series} for the times at infinity and the lapse at the puncture ($\alpha^{-}\mid_{x=0}$), the throat ($\alpha^{\pm}\mid_{r_{C}}$), and the event horizon ($\alpha^{\pm}\mid_{r_{+}}$) can now be reduced to expanding $C$ and the integrals $H_{C}(\infty)$, $K_{C}(r_{+})$, and $K_{C}(\infty)$ in $\delta$. 
This one can see from the formulas for the $\tau$'s and $\alpha$'s, (\ref{eq:tauevenfinal}), (\ref{eq:tauzgpfinal}) and (\ref{eq:alphaevenfinal}), (\ref{eq:alphazgpfinal}), for the even and ``zgp'' boundary conditions.

\subsection{Expanding the slice label $C(\delta)$}
With $r_{C}$ the unique (double counting for \hbox{$C = C_{lim}$}) root of the polynomial $p_{C}(r)$, it follows that 
\bea
\label{eq:Cofdproof}
	C & = & r_{C} \sqrt{- r_{C}^2 + 2M r_{C} - qM^{2}} \nonumber \\
	  & = & r_{C} \sqrt{r_{+} - r_{C}} 
		      \sqrt{r_{C} - r_{-}} \nonumber \\
	  & = & (\delta + r_{C_{lim}}) 
		\sqrt{\sigma_{+} - \delta} 
		\sqrt{\delta - \sigma_{-}}. 
\eea  
Here (\ref{eq:defdeltaRN}) has been used and without loss of generality the positive root has been chosen to yield \hbox{$0 \leq C \leq C_{lim}$}, since the negative root, \hbox{$- C_{lim} \leq C \leq 0$}, together with an inversion of the time direction, \hbox{$\tau \to - \tau$}, would result in the same maximal foliation.  
Making use of the positive constant $\mu$ defined in (\ref{eq:MU}) and expanding $C$ up to second order in $\delta$ yields
\beq
\label{eq:Cseries}
	C = C_{lim} + \frac{M - \mu}
			   {2\sqrt{ - \sigma_{+} \sigma_{-}}}\:\delta^{2} + {\cal O} (\delta^{3}).
\eeq
Here \hbox{$C_{lim} = r_{C_{lim}} \sqrt{- \sigma_{+} \sigma_{-}}$} can be read off from (\ref{eq:Cofdproof}) in the limit \hbox{$\delta \to 0$}. 
For later use (\ref{eq:Cseries}) is differentiated with respect to $\delta$ to obtain
\beq
\label{eq:dCddproof}
	\frac{dC}{d\delta} =  \frac{M - \mu}{\sqrt{- \sigma_{+} \sigma_{-}}}\:\delta
			      + {\cal O} (\delta^{2}),
\eeq
an upper bound for \hbox{$C_{lim} - C$} is given by 
\beq 
\label{eq:ClimminusCbound}
	0 \leq C_{lim} - C
	  \leq \frac{C_{lim}}{\delta^{2}\mid_{C=0}} \delta^{2} 
	  = \frac{C_{lim}}{\sigma_{+}^{2}} \delta^{2} ,
\eeq
and also the difference between $C_{lim}^{2}$ and $C^2$ is calculated,
\beq
\label{eq:Clim2minusC2}
  	C_{lim}^2 - C^{2} = \delta^{2}(\delta^{2} + \mu \delta + \nu),
\eeq
using $\nu$ as in (\ref{eq:NU}).

\subsection{Expanding the time at infinity $\tau(\delta)$}

\subsubsection{First Task}
In order to prove the statements for the exponential decay of $\delta$ with time at infinity, one has to show for even boundary conditions that 
\beq
\label{eq:Htoproof}
	\tau_{even}  = H_{C}(\infty) = -\Omega \ln{\left[ \delta \right]} + \Lambda + {\cal O} (\delta)
\eeq
holds, which inverted with the correct expressions for $\Omega$ and $\Lambda$ as in (\ref{eq:Omegais}) and (\ref{eq:Lambdais}) implies (\ref{eq:deltaevenRN}). 
Once this is shown for the even case, together with (\ref{eq:Cofdproof}) the result for the ``zgp'' case follows trivially from (\ref{eq:tauzgpfinal}) since
\beq
	\tau_{zgp}  = 2 H_{C}(\infty) - \frac{C}{M} 
		    =  - 2\Omega \ln{\left[ \delta \right]}
		         + (2\Lambda - \frac{C_{lim}}{M}) 
			 +  {\cal O} (\delta). 
\eeq  
The basic feature of the proof now consists of finding approximations $\cal H$ for $H_{C}(\infty)$ which, being straightforward to integrate and to expand in $\delta$, allow estimates for the remainder in order to complete the task of showing (\ref{eq:Htoproof}). 
The necessary preparations and estimates will be derived in the following.

\subsubsection{Preparations}
We start with
\beq
\label{eq:Hproof} 
	H_{C}(\infty) = - \int\limits^{\infty}_{\delta} \frac{C(\delta) \: ds}{f(r) \sqrt{p_{C}(r)}} 
\eeq
obtained by substituting 
\beq
\label{eq:syrlim}
	r = s + r_{C_{lim}}
\eeq
in the integral (\ref{eq:Hintegral}) to be calculated for the time at infinity (\ref{eq:tauevenfinal}), and write $C(\delta)$ as in (\ref{eq:Cofdproof}).
For the term $\frac{1}{f(r)}$ one can then find with \hbox{$\lambda_{\pm} \:^{>}_{<} 0$} as in (\ref{eq:LAMBDA}) the following expansion into a partial fraction, 
\bea 
\label{eq:onefpartialfraction} 
	\frac{1}{f(r)} & = & \frac{(s+r_{C_{lim}})^{2}}
                                              {(s+r_{C_{lim}})^{2} - 2M(s+r_{C_{lim}}) + qM^{2}} \nonumber \\
      			           & = & 1 + \frac{\lambda_{+}}{s - \sigma_{+}} + \frac{\lambda_{-}}{s- \sigma_{-}}.	
\eea
Also one should observe that for the polynomial $p_{C_{lim}}(r)$ corresponding to the limiting value $C_{lim}$ with $r_{C_{lim}}$ being a double root,
\bea
	p_{C_{lim}}(r) & = & r^{4}f(r) + C_{lim}^2 \\ \nonumber
		       & = & (r-r_{C_{lim}})^{2} \: (r^2 + 2(r_{C_{lim}}-M)r + \frac{C_{lim}^{2}}{r_{C_{lim}}^{2}}),
\eea
one can by substituting (\ref{eq:syrlim}) write
\beq
	p_{C_{lim}}(r) = s^{2}(s^{2} + \mu s + \nu).
\eeq
Hence together with (\ref{eq:Clim2minusC2}) we obtain the expression
\bea 
	p_C(r) & = & p_{C_{lim}}(r) - C_{lim}^2 + C^{2} \nonumber \\
	       & = & s^{2}(s^{2} + \mu s + \nu) - \delta^{2}(\delta^{2} + \mu \delta + \nu). \ \ 	
\eea
With \hbox{$\mu, \nu > 0$} the inequalities 
\bea
\label{eq:polydown}
	0 \leq \nu (s^{2} - \delta^{2}) & \leq & p_{C}(r) \\ 
\label{eq:polyup}
					     p_{C}(r) & \leq & (s^{2} - \delta^{2}) 
								\left( \nu + 2(s^{2} + \mu s) \right) \ \ 
\eea
hold for \hbox{$0 \leq \delta \leq s$}, which allow us to derive the two basic estimates needed in what follows.
The first,
\bea 
\label{eq:estimate1}		
	\bigg| \frac{1}{\sqrt{p_{C}(r)}} 
	   & - & \frac{1}{\sqrt{\nu(s^{2} - \delta^{2})}} \bigg| \nonumber \\
	   & = & \frac{\sqrt{p_{C}(r)} - \sqrt{\nu(s^{2} - \delta^{2})}}
                      {\sqrt{\nu(s^{2} - \delta^{2})\: p_{C}(r)}} \nonumber \\
	   & \leq & \frac{\sqrt{\nu + 2(s^{2} + \mu s)} - \sqrt{\nu}}
			 {\sqrt{\nu\:p_{C}(r)}} \nonumber \\
           & \leq & \frac{\sqrt{1+\frac{2}{\nu} (s^{2} + \mu s)} - 1}
			 {\sqrt{\nu(s^{2} - \delta^{2})}} \nonumber \\
	   & \leq & \frac{s^{2} + \mu s}{\nu^{\frac{3}{2}} \sqrt{s^{2} - \delta^{2}} } ,
\eea
is found by applying (\ref{eq:polyup}) in the numerator, (\ref{eq:polydown}) in the denominator and using \hbox{$\sqrt{1 + x} \leq 1 + \frac{x}{2},\ x \geq 0$}. 
With 
\beq
	\frac{d}{dC} \left[ \frac{C}{\sqrt{p_{C}(r)}} \right] 
      = \frac{r^{4}f(r)}{p_{C}(r)^{\frac{3}{2}}}
\eeq
the mean value theorem can be applied to yield together with (\ref{eq:ClimminusCbound}) the inequality
\beq 
	\left| \frac{1}{f(r)} \left( \frac{C}{\sqrt{p_{C}(r)}} 
	             - \frac{C_{lim}}{\sqrt{p_{C_{lim}}(r)}} \right) \right| 
          	\leq \frac{C_{lim}}{\sigma_{+}^{2}} \delta^{2}
		     \frac{r^{4}}{p_{C}(r)^{\frac{3}{2}}}, 
\eeq
which by continuity is also valid for \hbox{$r = r_{+}$}. 
Using (\ref{eq:syrlim}), and also (\ref{eq:polydown}), this can be written in the form
\bea 
\label{eq:estimate2}
        && \bigg| \frac{1}{f(r)}
                \bigg( \frac{C(\delta)}
	                     {\sqrt{p_{C}(r)}}
	             -  \frac{C_{lim}}
                             {\sqrt{p_{C_{lim}}(r)}}
                \bigg) \bigg| \nonumber \\
	&& \ \ \ \ \ \ \ \ \ \ \ \ \ \ \ \ \ \ \ \leq \frac{C_{lim}}{\sigma_{+}^{2}} \delta^{2} 
	         \frac{(s + r_{C_{lim}})^{4}}{p_{C}(r)^{\frac{3}{2}}} \nonumber \\
 	&& \ \ \ \ \ \ \ \ \ \ \ \ \ \ \ \ \ \ \ \leq \frac{C_{lim}}{\sigma_{+}^{2}} \delta^{2} 
	     \frac{(s + r_{C_{lim}})^{4}}{(\nu (s^{2} - \delta^{2}))^{\frac{3}{2}}}. 
\eea
It is worth mentioning that for the two estimates (\ref{eq:estimate1}) and (\ref{eq:estimate2}) contact with Eq.~(3.12) and (3.13) in \cite{Beig98} can be made by recalling the values \hbox{$^{0}\mu = 4M$} and \hbox{$^{0}\nu = \frac{9}{2} M^{2}$} for zero charge.

\subsubsection{Estimate for the integral $H_{C}(\infty)$}
The integration domain for the integration taking place in (\ref{eq:Hproof}) is split into \hbox{$A: 0 < \delta \leq s \leq \sqrt{\delta \sigma_{+}} < \sigma_{+}$} and \hbox{$B: \sqrt{\delta \sigma_{+}} \leq s \leq  \infty$}.	
So we can write
\bea 
	H_{C}(\infty) & = & H^{A} + H^{B} \nonumber \\
	              & = & (H^{A} - {\cal H}^{A}) + {\cal H}^{A}
 		            + (H^{B} - {\cal H}^{B}) + {\cal H}^{B} \ \ \ \ \ \ 
\eea
introducing the approximation integrals ${\cal H}^{A}$ and ${\cal H}^{B}$ given by 
\bea 
	{\cal H}^{A} & = & - \int\limits^{\sqrt{\delta \sigma_{+}}}_{\delta} 
		      \frac{C_{lim} \:ds}{f(r) \sqrt{\nu(s^{2} - \delta^{2})}} , \\
	{\cal H}^{B} & = & - \int\limits_{\sqrt{\delta \sigma_{+}}}^{\infty} 
		      \frac{C_{lim} \:ds}{f(r) \sqrt{p_{C_{lim}}(r)}}. \ \ \ \ \ \ 
\eea
Using (\ref{eq:onefpartialfraction}) we obtain $\cal H^{A}$ as
\beq
	{\cal H}^{A} = - \frac{C_{lim}}{\sqrt{\nu}}
		    \left(
			{\cal I}^{A\star}+ \lambda_{+} {\cal I}^{A+} + \lambda_{-} {\cal I}^{A-} 
		    \right),
\eeq
where the following integrals can be computed explicitly (using e.g.\ Chap.~21 of \cite{Bronstein97}) and expanded in $\delta$ in the late time limit \hbox{$\delta \to 0$} as
\bea
		{\cal I}^{A\star} & = & \int\limits^{\sqrt{\delta \sigma_{+}}}_{\delta} 
		      		        \frac{ds}{\sqrt{s^{2} - \delta^{2}}} \\ 
			         & = & - \frac{1}{2} \ln{\left[ \delta  \right]}
				    + \ln{\left[ 2\sqrt{\sigma_{+}} \right]} 
				    + {\cal O} (\delta) \nonumber , \\
		{\cal I}^{A\pm} & = & \int\limits^{\sqrt{\delta \sigma_{+}}}_{\delta} 
				    \frac{ds}{(s - \sigma_{\pm}) \sqrt{s^{2} - \delta^{2}}} \:ds \\
			        & = & \frac{1}{2\sigma_{\pm}} \ln{\left[ \delta  \right]}
				    - \frac{\ln{\left[ 2\sqrt{\sigma_{+}} \right]}}{\sigma_{\pm}} 
				    + o (1) , \nonumber
\eea
to yield
\bea 
\label{eq:HA}
    {\cal H}^{A} & = & \frac{C_{lim}}{2\sqrt{\nu}}
		           \left( 1 - (\frac{\lambda_{+}}{\sigma_{+}} 
                                    + \frac{\lambda_{-}}{\sigma_{-}}) \right) 
                           \ln{\left[ \delta \right]} \\  
                 & \ & - \frac{C_{lim}}{\sqrt{\nu}} 
			   \left( 1 - (\frac{\lambda_{+}}{\sigma_{+}} 
                                    + \frac{\lambda_{-}}{\sigma_{-}}) \right)
			   \ln{\left[ 2\sqrt{\sigma_{+}} \right]} + o (1). \nonumber 
\eea
Similarly, for $\cal H^{B}$ written as
\beq
	{\cal H}^{B} = - C_{lim}
		    \left(
			{\cal I}^{B\star} + \lambda_{+} {\cal I}^{B+} + \lambda_{-} {\cal I}^{B-} 
		    \right) ,
\eeq
when introducing \hbox{$\xi_{\pm} \geq 0$} as in (\ref{eq:XI}), with
\bea
		{\cal I}^{B\star} & = & \int\limits_{\sqrt{\delta \sigma_{+}}}^{\infty} 
		  		    \frac{ds}{\sqrt{p_{C_{lim}}(r)}} \\
			      & = & - \frac{1}{2\sqrt{\nu}}
					\ln{\left[ \delta \right]
				    - \frac{1}{\sqrt{\nu}} \ln{
					\left[
					  \frac{2\sqrt{\nu}+\mu}{4\nu} \sqrt{\sigma_{+}}
					\right]}}
                                    + o (1) , \nonumber \\
		{\cal I}^{B\pm} & = & \int\limits_{\sqrt{\delta \sigma_{+}}}^{\infty} 
				    \frac{ds}{(s - \sigma_{\pm})\sqrt{p_{C_{lim}}(r)}} \:ds \\
			      & = & \frac{1}{2\sqrt{\nu}\sigma_{\pm}}
					\ln{\left[ \delta \right] 
				    + \frac{1}{\sqrt{\nu}\sigma_{\pm}} \ln{
					\left[
				 	  \frac{2\sqrt{\nu}+\mu}{4\nu} \sqrt{\sigma_{+}}
					\right]}} \nonumber \\
			      & \ & + \frac{\xi_{\pm}}{\lambda_{\pm}} + o (1) , \nonumber
\eea
we obtain
\bea 
\label{eq:HB}
	{\cal H}^{B} & = & \frac{C_{lim}}{2\sqrt{\nu}}
		               \left( 1 - (\frac{\lambda_{+}}{\sigma_{+}} 
                                         + \frac{\lambda_{-}}{\sigma_{-}}) \right) 
                               \ln{\left[ \delta \right]} \\
                     & \ & + \frac{C_{lim}}{\sqrt{\nu}} 
			       \left( 1 - (\frac{\lambda_{+}}{\sigma_{+}} 
                                         + \frac{\lambda_{-}}{\sigma_{-}}) \right) 
			       \ln{\left[ \frac{2\sqrt{\nu}+\mu}{4\nu} \sqrt{\sigma_{+}} \right]} \nonumber \\
		     & \ & - C_{lim}(\xi_{+} + \xi_{-}) + o (1). \nonumber
\eea
Estimating the remainders for \hbox{$\triangle H^{A} = \left| H^{A} - {\cal H}^{A} \right|$} now yields
\bea 
\label{eq:estiHA}
	\triangle H^{A}
	   & = & \Bigg|
		    \int\limits^{\sqrt{\delta \sigma_{+}}}_{\delta}	
		    \frac{1}{f(r)}  
                    \Bigg(
			\frac{C(\delta)}{\sqrt{p_{C}(r)}} - \frac{C_{lim}}{\sqrt{\nu(s^{2} - \delta^{2})}} 
                    \Bigg) ds
	         \Bigg| \nonumber \\
	 & \leq & \Bigg|
		    \int\limits^{\sqrt{\delta \sigma_{+}}}_{\delta}	
		    \frac{1}{f(r)} 
		    \frac{C(\delta) - C_{lim}}{\sqrt{p_{C}(r)}}\:ds
	         \Bigg| \nonumber \\
	   & + & \Bigg|
		    \int\limits^{\sqrt{\delta \sigma_{+}}}_{\delta}	
		    \frac{C_{lim}}{f(r)} 
           	      \Bigg(
			\frac{1}{\sqrt{p_{C}(r)}} - \frac{1}{\sqrt{\nu(s^{2} - \delta^{2})}} 
                      \Bigg)ds
	         \Bigg| \nonumber \\
	 & \leq & \begin{rm}{const}\end{rm} \cdot \int\limits^{\sqrt{\delta \sigma_{+}}}_{\delta} 
			      \frac{s\:ds}{\sqrt{s^{2} - \delta^{2}}}
	   = {\cal O} (\sqrt{\delta}). 			
\eea
Here (\ref{eq:Cseries}) has been used to neglect the first and (\ref{eq:estimate1}) to bound the second summand.
Given the second basic estimate (\ref{eq:estimate2}), \hbox{$\triangle H^{B} = \left| H^{B} - {\cal H}^{B} \right|$} in the limit \hbox{$\delta \to 0$} has a bound of the form
\bea 
	\triangle H^{B} 
	   & = & \Bigg|
		    \int\limits_{\sqrt{\delta \sigma_{+}}}^{\infty}	
		    \frac{1}{f(r)} 
	   	     \Bigg(
			\frac{C(\delta)}{\sqrt{p_{C}(r)}} - \frac{C_{lim}}{\sqrt{p_{C_{lim}}(r)}} 
                     \Bigg) ds
	         \Bigg| \ \nonumber \\
	 & \leq & \begin{rm}{const}\end{rm} \cdot \delta^{2} \:
		    \int\limits_{\sqrt{\delta \sigma_{+}}}^{\infty}	
		    \frac{(s+r_{C_{lim}})^{4}\:ds}{(s^{2}(s^{2} + \mu s + \nu))^{\frac{3}{2}}}.  		
\eea
Following \cite{Beig98} and splitting this integral into
\bea
	\int\limits_{\sqrt{\delta \sigma_{+}}}^{1}	
		\frac{(s+r_{C_{lim}})^{4}\:ds}{(s^{2}(s^{2} + \mu s + \nu))^{\frac{3}{2}}}	
        & \leq & \begin{rm}{const}\end{rm} \cdot \int\limits_{\sqrt{\delta \sigma_{+}}}^{1}	
		\frac{ds}{(s^{2} - \delta^{2})^{\frac{3}{2}}} \nonumber \\
        &   =  & {\cal O} (\frac{1}{\delta})        
\eea
and
\beq
	\int\limits^{\infty}_{1}	
		\frac{(s+r_{C_{lim}})^{4}\:ds}{(s^{2}(s^{2} + \mu s + \nu))^{\frac{3}{2}}}	
        \leq \begin{rm}{const}\end{rm} \cdot \int\limits^{\infty}_{1} \frac{ds}{s^{2}} 
        = {\cal O} (1), 
\eeq
we finally find
\beq
\label{eq:estiHB}
	\left| H^{B} - {\cal H}^{B} \right| = {\cal O} (\delta). 
\eeq
Putting the results for the ${\cal H}$'s and the estimates of the remainders \hbox{$\triangle H = \left| H - {\cal H} \right|$}, (\ref{eq:HA},\ \ref{eq:HB}) and (\ref{eq:estiHA},\ \ref{eq:estiHB}), together we obtain 
\beq
\label{eq:Hinftynotgoodenough}
 	H_{C}(\infty) = -\Omega \ln{\left[ \delta \right]} + \Lambda + o (1)	
\eeq
with $\Omega$ and $\Lambda$ as proposed in (\ref{eq:Omegais}) and (\ref{eq:Lambdais}). 
However, comparing (\ref{eq:Hinftynotgoodenough}) with the original task as formulated in (\ref{eq:Htoproof}), it turns out that our estimates so far are (as in \cite{Beig98}) ``not quite good enough'' to conclude the form of the higher order term appearing in the exponential decay of $\delta$ with $\tau$. 
The estimates give $\Omega$ and $\Lambda$, but we now proceed to improve the estimate for the $o(1)$ term.

\subsubsection{Second task}
Next, we have to show that in the limit \hbox{$\delta \to 0$} the last equality in
\beq
\label{eq:Ktoproof}
	\frac{dH_{C}(\infty)}{d\delta} 
 		= - \frac{1}{2} K_{C}(\infty) \frac{dC}{d\delta} 
	        = - \Omega \frac{1}{\delta} + {\cal O} (1)
\eeq
holds.
This expression is obtained by making contact between $H_{C}(\infty)$ and $K_{C}(\infty)$ using (\ref{eq:HisKhalf}).
Note that integrating (\ref{eq:Ktoproof}), which yields
\beq
	H_{C}(\infty) = - \Omega \ln{\left[ \delta  \right]} + \Lambda^{*} + {\cal O} (\delta),
\eeq
and inferring \hbox{$\Lambda = \Lambda^{*}$} finally proves the original task formulated in (\ref{eq:Htoproof}). 
So with (\ref{eq:dCddproof}) one has to show
\beq
\label{eq:Kinfseries}
	K_{C}(\infty) = \frac{2\Omega\sqrt{-\sigma_{+}\sigma_{-}}}{M-\mu}\:\frac{1}{\delta^{2}} 
                  + {\cal O} (\frac{1}{\delta}).	
\eeq

\subsubsection{Estimate for the integrals $K_{C}(r_{+})$ and $K_{C}(\infty)$}
Substituting (\ref{eq:syrlim}) in (\ref{eq:Kintegral}), we can write $K_{C}(\infty)$ as
\beq
\label{eq:Kproof}
	K_{C}(\infty) = \int\limits^{\infty}_{\delta}
			\left(
			   1 + \frac{\chi_{+}}{(s - \frac{M - \mu}{2})^{2}} 
                             + \frac{\chi_{-}}{s^{2}}
		        \right)
			\frac{ds}{\sqrt{p_{C}(r)}},
\eeq
making use of a partial fraction with $\chi_{\pm}$ defined in (\ref{eq:CHI}).
For $K_{C}(\infty)$ the integration domain can be split at the event horizon, which is appropriate since one is later on also interested in $K_{C}(r_{+})$.
Note that this was not possible for $H_{C}(r)$ since this integral diverges at $r_{+}$.
Hence \hbox{$A: 0 < \delta \leq s \leq \sigma_{+}$} and \hbox{$B: \sigma_{+} \leq s \leq \infty$} are the integration domains. 	
Again we can write
\bea
	K_{C}(\infty) & = & K^{A} + K^{B} \nonumber \\ 
	              & = & (K^{A} - {\cal K}^{A}) + {\cal K}^{A}
 		            + (K^{B} - {\cal K}^{B}) + {\cal K}^{B} \ \ \ \ \ \ 
\eea
and introduce approximation integrals ${\cal K}^{A}$ and ${\cal K}^{B}$ by
\bea
	{\cal K}^{A} & = & \int\limits_{\delta}^{\sigma_{+}}
			   \left(
			         1 + \frac{\chi_{+}}{(s - \frac{M - \mu}{2})^{2}} 
                                   + \frac{\chi_{-}}{s^{2}}
		           \right) 
		           \frac{ds}{\sqrt{\nu(s^{2} - \delta^{2})}} \nonumber \\
                     & = & {\cal J}^{A\star}+ \chi_{+} {\cal J}^{A+} + \chi_{-} {\cal J}^{A-} , \\ 
	{\cal K}^{B} & = & \int\limits_{\sigma_{+}}^{\infty}
			   \left(
			         1 + \frac{\chi_{+}}{(s - \frac{M - \mu}{2})^{2}} 
                                   + \frac{\chi_{-}}{s^{2}}
		           \right) 
		           \frac{ds}{\sqrt{p_{C_{lim}}(r)}} \nonumber \\
                     & = & {\cal J}^{B\star}+ \chi_{+} {\cal J}^{B+} + \chi_{-} {\cal J}^{B-}.
\eea

Explicitly computing and expanding the $\cal J$'s yields
\bea
	{\cal J}^{A\star} & = & \int\limits_{\delta}^{\sigma_{+}}
		          	 \frac{ds}{\sqrt{\nu(s^{2} - \delta^{2})}} 
			   =  {\cal O} (\ln{\left[ \delta \right]}) , \\ 
	{\cal J}^{A+}     & = & \int\limits_{\delta}^{\sigma_{+}}
			   	\frac{ds}{(s - \frac{M - \mu}{2})^{2} \sqrt{\nu(s^{2} - \delta^{2})}} 
			   =  {\cal O} (\ln{\left[ \delta \right]}) , \ \ \ \ \\ 
	{\cal J}^{A-}     & = & \int\limits_{\delta}^{\sigma_{+}}
			   	\frac{ds}{s^{2} \sqrt{\nu(s^{2} - \delta^{2})}} 
			   =  \frac{1}{\sqrt{\nu}} \frac{1}{\delta^{2}} + {\cal O} (1).  
\eea
Abbreviating \hbox{$\ln{\left[ \cdots \right]} = \ln{\left[
					   \frac{\sigma_{+}\mu + 2(\nu+\sqrt{\nu}
                                               \sqrt{\sigma_{+}^{2} + \sigma_{+}\mu + \nu})}
						{\sigma_{+}(\mu+2\sqrt{\nu})}
					\right]}$} 
also
\bea
		{\cal J}^{B\star} & = & \int\limits_{\sigma_{+}}^{\infty} 
		  		  	  \frac{ds}{\sqrt{p_{C_{lim}}(r)}} 
		    	           =  \frac{1}{\sqrt{\nu}} \ln{\left[ \cdots \right]} , \\ 
		{\cal J}^{B+} & = & \int\limits_{\sigma_{+}}^{\infty} 
				    \frac{ds}{(s-\frac{M - \mu}{2})^{2}	\sqrt{p_{C_{lim}}(r)}} \\ \nonumber
			      & = & \frac{16(2\sigma_{-} - M +2\sqrt{\sigma_{+}^{2} + \sigma_{+}\mu + \nu})}
                                         {(2\sigma_{-} - M)^{2}(M-\mu)^{2}} \\ \nonumber
			      & \ & \ \ + \frac{1}{\sqrt{\nu}}
                                    \frac{4}{(M - \mu)^{2}} 
                                    \ln{\left[ \cdots \right]} , \\ 
		{\cal J}^{B-} & = & \int\limits_{\sigma_{+}}^{\infty} 
				    \frac{ds}{s^{2} \sqrt{p_{C_{lim}}(r)}} \\ \nonumber
			      & = & \frac{3\sigma_{+}^{2}\mu+(2\nu-3\sigma_{+}\mu)
						\sqrt{\sigma_{+}^{2} + \sigma_{+}\mu + \nu}}
			        	 {(2\sigma_{+}\nu)^{2}} \\ \nonumber 
			       & \ & \ \ +
				    \frac{1}{\sqrt{\nu}} 
                                    \frac{3\mu^{2} - 4\nu}{8\nu^{2}} 
                                    \ln{\left[ \cdots \right]}
\eea
are found.
Summarizing results, 
\beq
\label{eq:KA}
	{\cal K}^{A} = \frac{\chi_{-}}{\sqrt{\nu}} \frac{1}{\delta^{2}} + {\cal O} (\ln{\left[ \delta \right]}) 
\eeq
and with $\eta$ defined already in (\ref{eq:ETA}) we obtain
\beq
\label{eq:KB}
        {\cal K}^{B} = {\cal J}^{B\star} + \chi_{+} {\cal J}^{B+} + \chi_{-} {\cal J}^{B-}
                     = \eta = {\cal O} (1) .
\eeq
Last, but not least, the remainders in the limit \hbox{$\delta \to 0$} can be estimated for \hbox{$\triangle K^{A} = \left| K^{A} - {\cal K}^{A} \right|$}, 
\bea 
\label{eq:estiKA}
	\triangle K^{A}
                  & = & \Bigg|
			\int\limits_{\delta}^{\sigma_{+}}
			\left(
			   1 + \frac{\chi_{+}}{(s - \frac{M - \mu}{2})^{2}} 
                             + \frac{\chi_{-}}{s^{2}}
		        \right) \nonumber \\
		   & \ & \ \ \ \ \left(
			\frac{1}{\sqrt{p_{C}(r)}}
			- \frac{1}{\sqrt{\nu(s^{2} - \delta^{2})}}
			\right) ds
                        \Bigg| \nonumber \\
               & \leq & \begin{rm}{const}\end{rm} \cdot
			\int\limits_{\delta}^{\sigma_{+}} \frac{s\:ds}{\sqrt{s^2-\delta^2}}
		   =  {\cal O} (\frac{1}{\delta}),			
\eea
using (\ref{eq:estimate1}) and for \hbox{$\triangle K^{B} = \left| K^{B} - {\cal K}^{B} \right|$}, 
\bea
\label{eq:estiKB}
	\triangle K^{B} 
                  & = & \Bigg|
			\int\limits_{\sigma_{+}}^{\infty}
			\left(
			   1 + \frac{\chi_{+}}{(s - \frac{M - \mu}{2})^{2}} 
                             + \frac{\chi_{-}}{s^{2}}
		        \right) \nonumber \\
	          & \ & \ \ \ \ \left(
			\frac{1}{\sqrt{p_{C}(r)}} 
			- \frac{1}{\sqrt{p_{C_{lim}}(r)}}
			\right) ds
                        \Bigg| \nonumber \\
	       & \leq & \begin{rm}{const}\end{rm} \cdot \delta^{2} \:
                        \int\limits_{\sigma_{+}}^{\infty}
			\frac{f(r) (s+r_{C_{lim}})^{4} \:ds}
                             {(s^{2}(s^{2} + \mu s + \nu))^{\frac{3}{2}}} 
                   =  {\cal O} (\delta^{2}), \ \ \ \ \ \ \ 	
\eea 
using (\ref{eq:estimate2}) while considering (\ref{eq:Cseries}).
Now taking finally into account the results for the ${\cal K}$'s and the estimates \hbox{$\triangle K = \left| K - {\cal K} \right|$}, (\ref{eq:KA},\ \ref{eq:KB}) and (\ref{eq:estiKA},\ \ref{eq:estiKB}), together with
\beq
	\frac{\chi_{-}}{\sqrt{\nu}} = \frac{2\Omega\sqrt{-\sigma_{+}\sigma_{-}}}{M-\mu}
\eeq
the statements (\ref{eq:Kinfseries}) and therefore (\ref{eq:Htoproof}) are proven. 
Furthermore, since $K_{C}(\infty)$ and $K_{C}(r_{+})$ differ by $K^{B}$ only, we can conclude from (\ref{eq:KB}) and (\ref{eq:estiKB}) that
\beq
\label{eq:KinfminusKplusseries}
	K_{C}(\infty) - K_{C}(r_{+}) = \eta + {\cal O} (\delta^{2})
\eeq
holds.
We can use this expression to extend the study of \cite{Beig98} since (\ref{eq:KinfminusKplusseries}) allows us to obtain expansions in $\delta$ not only at the throat but also at the event horizon.  

\subsection{Expanding the multiplicator function $\Phi(\delta)$ and the lapse $\alpha(\delta)$}
Using (\ref{eq:Kinfseries}) the multiplicator function $\Phi$ can be expanded in $\delta$ as
\beq
\label{eq:Useries}
	\Phi = \frac{1}{2}\:\frac{K_{C}(\infty)}{K_{C}(\infty) + \frac{1}{M}}
             = \frac{1}{2} - \frac{\sqrt{\nu}}{2\chi_{-}M}\:\delta^{2} +  {\cal O} (\delta^{3}).
\eeq
So at late times the ``zgp'' lapse is in leading order the average of the odd and even lapse.

The lapse at the puncture is minus and plus one for the odd and even boundary conditions, respectively, and therefore at \hbox{$x = 0$} the ``zgp'' lapse constructed as a linear combination is obtained as
\bea
\label{eq:firstalpha}
        \alpha^{-}_{zgp} \mid _{x=0} & = & \Phi \cdot \alpha^{-}_{even} \mid _{x=0} 
                                        + (1-\Phi) \cdot \alpha^{-}_{odd} \mid _{x=0} \nonumber \\
				     & = & 2\Phi-1 
                                     = - \frac{\sqrt{\nu}}{\chi_{-}M}\:\delta^{2} +  {\cal O} (\delta^{3}). 
\eea

Next, the value of the lapse at the throat $r_{C}$ is found for the even case from (\ref{eq:alphaevenfinal}) using (\ref{eq:defdeltaRN}),
\bea
	\alpha^{\pm}_{even} \mid _{r_{C}} 
                                    & = & - \frac{1}{K_{C}(\infty)} \:
                                            \frac{\delta+r_{C_{lim}}}
                                                 {\delta(\delta-\frac{M - \mu}{2})} \nonumber \\
 				    & = & \frac{2r_{C_{lim}}\sqrt{\nu}}{\chi_{-}(M - \mu)}\:\delta +  {\cal O} (\delta^{2}) ,
\eea
and for the ``zgp'' boundary conditions it is at leading order half of this value,
\bea
	\alpha^{\pm}_{zgp} \mid _{r_{C}} & = & \Phi \cdot \alpha^{\pm}_{even} \mid _{r_{C}} \nonumber \\
                                         & = & \frac{r_{C_{lim}}\sqrt{\nu}}{\chi_{-}(M - \mu)}\:\delta 
						+  {\cal O} (\delta^{2}),
\eea
since the odd lapse vanishes at the throat and does not contribute in the late time average.

Finally, the lapse at the event horizon can be expanded in $\delta$. Remembering that $r_{+}$ is a root of $f(r)$, for the odd lapse it follows that \hbox{$\alpha^{\pm}_{odd} \mid _{r_{+}} = \pm \frac{C}{r_{+}^{2}}$}.
The even lapse at $r_{+}$ is given by
\bea
	\alpha^{\pm}_{even} \mid _{r_{+}} 
		& = & -\frac{1}{K_{C}(\infty)} 
		       \left( \frac{4}{\sigma_{+} - \sigma_{-}} 
                            - \frac{C}{r_{+}^{2}} K_{C}(r_{+})  \right) \nonumber \\	
	        & = & \frac{C_{lim}}{r_{+}^{2}} +  
		      \frac{\sqrt{\nu}}{\chi_{-}}
		      \Bigg(
			-\frac{4}{\sigma_{+}-\sigma_{-}}
			-\frac{1}{r_{+}^{2}}
			(C_{lim}\eta \nonumber \\
		& \ & \ \ \
			- \frac{M - \mu}{2\sqrt{-\sigma_{+}\sigma_{-}}}\frac{\chi_{-}}{\sqrt{\nu}})
		      \Bigg) \:\delta^{2}
		      + {\cal O} (\delta^{3}) ,	 
\eea
where the expansions (\ref{eq:Cofdproof}), (\ref{eq:Kinfseries}), and (\ref{eq:KinfminusKplusseries}) have been used.
With the series (\ref{eq:Useries}) for $\Phi$ it turns out that for the ``zgp'' boundary conditions on the left-hand side of the throat the value of the lapse at the event horizon collapses to zero as
\bea
	\alpha_{zgp}^{-} \mid _{r_{+}} 
		& = & \Phi \cdot \alpha^{-}_{even} \mid _{r_{+}} 
                      + (1-\Phi) \cdot \alpha^{-}_{odd} \mid _{r_{+}} \nonumber	\\
	        & = & 
	            \frac{\sqrt{\nu}}{\chi_{-}}		  
		    \Bigg(
			-\frac{2}{\sigma_{+}-\sigma_{-}} \nonumber \\
                & \ & \ \ \ 
			-\frac{C_{lim}}{r_{+}^{2}}(\frac{1}{M} + \frac{\eta}{2})
		    \Bigg)\:\delta^{2}
		    + {\cal O} (\delta^{3}). \ \ \ \ 
\eea
On the right-hand side, though, with 
\bea
\label{eq:lastalpha}
	\alpha_{zgp}^{+} \mid _{r_{+}} 
		& = & \Phi \cdot \alpha^{+}_{even} \mid _{r_{+}} 
                      + (1-\Phi) \cdot \alpha^{+}_{odd} \mid _{r_{+}} \nonumber	\\
	        & = & \frac{C_{lim}}{r_{+}^{2}} +
	              \frac{\sqrt{\nu}}{\chi_{-}}
		      \Bigg(
			-\frac{2}{\sigma_{+}-\sigma_{-}}
		        -\frac{1}{2r_{+}^{2}}
                        (C_{lim}\eta \nonumber \\
		& \ & \ \ \ 
			- \frac{M-\mu}{\sqrt{-\sigma_{-}\sigma_{+}}}
			  \frac{\chi_{-}}{\sqrt{\nu}})
		      \Bigg)\:\delta^{2}
		    + {\cal O} (\delta^{3}) \ \ \ \ 
\eea
as for the even boundary conditions, the value $\frac{C_{lim}}{r_{+}^{2}}$ is approached.
The latter is an important example for the conjecture regarding the late time limit of the lapse at the right-hand event horizon stated in our previous paper \cite{mypaper1}.

\subsection{Comments}
We want to make several comments regarding the proof of the late time analysis.

Since the whole proof depends on the statements (\ref{eq:Htoproof}) and (\ref{eq:KinfminusKplusseries}) for $H_{C}(\infty)$ and \hbox{$K_{C}(\infty) - K_{C}(r_{+})$}, these two fundamental expressions have been checked numerically in \cite{mythesis}. 
This can be done by calculating for different values of $q$ the integrals $H_{C}(\infty)$, $K_{C}(r_{+})$, and $K_{C}(\infty)$ with Mathematica and by verifying that both \hbox{$\frac{1}{\delta} \left| H_{C}(\infty) - (-\Omega \ln{\left[ \delta \right]} + \Lambda) \right|$} and \hbox{$\frac{1}{\delta^{2}} \left| K_{C}(\infty) - K_{C}(r_{+}) - \eta \right|$} behave like ${\cal O} (1)$ in the limit \hbox{$\delta \to 0$}.
Hence not only the formulas (\ref{eq:Omegais},\ \ref{eq:Lambdais}) and (\ref{eq:ETA}) found for $\Omega, \Lambda$ and $\eta$ as functions of $q$ can be confirmed, but also the order of the ``rest term'' appearing in (\ref{eq:Htoproof}) and (\ref{eq:KinfminusKplusseries}) can be verified.

Note that astrophysically relevant black holes are expected to have \hbox{$0 \leq q \ll 1$} as it would be very difficult for any astrophysical body to achieve and/or maintain a charge to mass ratio \hbox{$\sqrt{q} = \frac{\mid Q \mid}{M}$} of greater than $10^{-18}$ \cite{Wald84}, since otherwise particles of opposite charge would be attracted selectively.
By analyzing the terms depending on $q$ appearing in the $\delta$ expansions, one can see that for moderate charges these expressions deviate only very little from their Schwarzschild values denoted by an upper index ``0'' referring to zero charge.
For quantitative details and plots of $\Omega$, the $\kappa$'s, and $\eta$ as a function of $q$ see \cite{mythesis}.

However, in the limit \hbox{$q \to 1$} the expressions for $\Omega$ and $\Lambda$ diverge like
\beq
	\Omega \stackrel{q \to 1}{\simeq} \frac{1}{\sqrt{1-q}}M
\eeq 
and
\beq
	\Lambda \stackrel{q \to 1}{\simeq} \frac{1}{2} \frac{\ln{\left[ 1-q \right]}}{\sqrt{1-q}}M 
					 + \frac{\ln{\left[ 2 \right]}}{\sqrt{1-q}}M. 	
\eeq
Furthermore, with $C_{lim}$ vanishing like
\beq
	C_{lim} \stackrel{q \to 1}{\simeq} \sqrt{1-q}M^{2}
\eeq
the pre-exponential factors approach zero for $q \to 1$ as
\beq
\label{eq:kappaone}
	\kappa_{even} \stackrel{q \to 1}{\simeq} \kappa_{zgp}  
		      \stackrel{q \to 1}{\simeq} 2\sqrt{1-q}M ,
\eeq
and the LTA obviously breaks down. 
This, however, is no surprise as for the extremal Reissner-Nordstr\"om spacetime containing a naked singularity our ansatz for $\delta$, (\ref{eq:defdeltaRN}), is not valid.

Using previously stated expressions, it should be straightforward for the reader interested in the full 4-metric at the event horizon to expand also the radial metric component, (\ref{eq:gammafinal}), and for even or ``zgp'' boundary conditions the shift, (\ref{eq:betafinal}). 
But note that both $\gamma$ and $\beta$ diverge at the throat.
This, however, is only a coordinate effect of the radial gauge. 
For numerical purposes it is therefore preferable to perform as in Subsecs.~\ref{subsec:LTAzeroeven} and \ref{subsec:LTAzerozgp} a late time expansion of the 3-metric (or the shift) in suitable spatial coordinates such as the isotropic (or, in the case of a non-vanishing shift, isothermal) grid coordinates.


\bibliographystyle{apsrev}

\bibliography{myreferences}

\begin{thebibliography}{30}
\expandafter\ifx\csname natexlab\endcsname\relax\def\natexlab#1{#1}\fi
\expandafter\ifx\csname bibnamefont\endcsname\relax
  \def\bibnamefont#1{#1}\fi
\expandafter\ifx\csname bibfnamefont\endcsname\relax
  \def\bibfnamefont#1{#1}\fi
\expandafter\ifx\csname citenamefont\endcsname\relax
  \def\citenamefont#1{#1}\fi
\expandafter\ifx\csname url\endcsname\relax
  \def\url#1{\texttt{#1}}\fi
\expandafter\ifx\csname urlprefix\endcsname\relax\def\urlprefix{URL }\fi
\providecommand{\bibinfo}[2]{#2}
\providecommand{\eprint}[2][]{\url{#2}}

\bibitem[{\citenamefont{York}(1979)}]{York79}
\bibinfo{author}{\bibfnamefont{J.}~\bibnamefont{York}}, in
  \emph{\bibinfo{booktitle}{Sources of Gravitational Radiation}}, edited by
  \bibinfo{editor}{\bibfnamefont{L.}~\bibnamefont{Smarr}}
  (\bibinfo{publisher}{Cambridge University Press},
  \bibinfo{address}{Cambridge, England}, \bibinfo{year}{1979}).

\bibitem[{\citenamefont{Estabrook et~al.}(1973)\citenamefont{Estabrook,
  Wahlquist, Christensen, DeWitt, Smarr, and Tsiang}}]{Estabrook73}
\bibinfo{author}{\bibfnamefont{F.}~\bibnamefont{Estabrook}},
  \bibinfo{author}{\bibfnamefont{H.}~\bibnamefont{Wahlquist}},
  \bibinfo{author}{\bibfnamefont{S.}~\bibnamefont{Christensen}},
  \bibinfo{author}{\bibfnamefont{B.}~\bibnamefont{DeWitt}},
  \bibinfo{author}{\bibfnamefont{L.}~\bibnamefont{Smarr}}, \bibnamefont{and}
  \bibinfo{author}{\bibfnamefont{E.}~\bibnamefont{Tsiang}},
  \bibinfo{journal}{Phys. Rev. D} \textbf{\bibinfo{volume}{7}},
  \bibinfo{pages}{2814 to 2817} (\bibinfo{year}{1973}).

\bibitem[{\citenamefont{Beig and Murchadha}(1998)}]{Beig98}
\bibinfo{author}{\bibfnamefont{R.}~\bibnamefont{Beig}} \bibnamefont{and}
  \bibinfo{author}{\bibfnamefont{N.~O.} \bibnamefont{Murchadha}},
  \bibinfo{journal}{Phys. Rev. D} \textbf{\bibinfo{volume}{57}},
  \bibinfo{pages}{4728 to 4737} (\bibinfo{year}{1998}),
  \bibinfo{note}{gr-qc/9706046}.

\bibitem[{\citenamefont{Eardley and Smarr}(1979)}]{Eardley79}
\bibinfo{author}{\bibfnamefont{D.}~\bibnamefont{Eardley}} \bibnamefont{and}
  \bibinfo{author}{\bibfnamefont{L.}~\bibnamefont{Smarr}},
  \bibinfo{journal}{Phys. Rev. D} \textbf{\bibinfo{volume}{19}},
  \bibinfo{pages}{2239 to 2259} (\bibinfo{year}{1979}).

\bibitem[{\citenamefont{Bernstein et~al.}(1989)\citenamefont{Bernstein, Hobill,
  and Smarr}}]{Bernstein89}
\bibinfo{author}{\bibfnamefont{D.}~\bibnamefont{Bernstein}},
  \bibinfo{author}{\bibfnamefont{D.}~\bibnamefont{Hobill}}, \bibnamefont{and}
  \bibinfo{author}{\bibfnamefont{L.}~\bibnamefont{Smarr}}, in
  \emph{\bibinfo{booktitle}{Frontiers in Numerical Relativity}}, edited by
  \bibinfo{editor}{\bibfnamefont{C.}~\bibnamefont{Evans}},
  \bibinfo{editor}{\bibfnamefont{L.}~\bibnamefont{Finn}}, \bibnamefont{and}
  \bibinfo{editor}{\bibfnamefont{D.}~\bibnamefont{Hobill}}
  (\bibinfo{publisher}{Cambridge University Press},
  \bibinfo{address}{Cambridge, England}, \bibinfo{year}{1989}), p.
  \bibinfo{pages}{57 to 73}.

\bibitem[{\citenamefont{Anninos
  et~al.}(1995{\natexlab{a}})\citenamefont{Anninos, Camarda, Mass{\'o}, Seidel,
  Suen, and Towns}}]{Anninos95}
\bibinfo{author}{\bibfnamefont{P.}~\bibnamefont{Anninos}},
  \bibinfo{author}{\bibfnamefont{K.}~\bibnamefont{Camarda}},
  \bibinfo{author}{\bibfnamefont{J.}~\bibnamefont{Mass{\'o}}},
  \bibinfo{author}{\bibfnamefont{E.}~\bibnamefont{Seidel}},
  \bibinfo{author}{\bibfnamefont{W.-M.} \bibnamefont{Suen}}, \bibnamefont{and}
  \bibinfo{author}{\bibfnamefont{J.}~\bibnamefont{Towns}},
  \bibinfo{journal}{Phys. Rev. D} \textbf{\bibinfo{volume}{52}},
  \bibinfo{pages}{2059 to 2082} (\bibinfo{year}{1995}{\natexlab{a}}),
  \bibinfo{note}{gr-qc/9503025}.

\bibitem[{\citenamefont{Br{\"u}gmann}(1999)}]{Bruegmann97}
\bibinfo{author}{\bibfnamefont{B.}~\bibnamefont{Br{\"u}gmann}},
  \bibinfo{journal}{Int. J. Mod. Phys. D} \textbf{\bibinfo{volume}{8}},
  \bibinfo{pages}{85} (\bibinfo{year}{1999}), \bibinfo{note}{gr-qc/9708035}.

\bibitem[{\citenamefont{Alcubierre
  et~al.}(2001{\natexlab{a}})\citenamefont{Alcubierre, Benger, Br\"ugmann,
  Lanfermann, Nerger, Seidel, and Takahashi}}]{Alcubierre00b}
\bibinfo{author}{\bibfnamefont{M.}~\bibnamefont{Alcubierre}},
  \bibinfo{author}{\bibfnamefont{W.}~\bibnamefont{Benger}},
  \bibinfo{author}{\bibfnamefont{B.}~\bibnamefont{Br\"ugmann}},
  \bibinfo{author}{\bibfnamefont{G.}~\bibnamefont{Lanfermann}},
  \bibinfo{author}{\bibfnamefont{L.}~\bibnamefont{Nerger}},
  \bibinfo{author}{\bibfnamefont{E.}~\bibnamefont{Seidel}}, \bibnamefont{and}
  \bibinfo{author}{\bibfnamefont{R.}~\bibnamefont{Takahashi}},
  \bibinfo{journal}{Phys. Rev. Lett.} \textbf{\bibinfo{volume}{87}},
  \bibinfo{pages}{271103} (\bibinfo{year}{2001}{\natexlab{a}}),
  \bibinfo{note}{gr-qc/0012079}.

\bibitem[{\citenamefont{Alcubierre et~al.}(2003)\citenamefont{Alcubierre,
  Br\"ugmann, Diener, Koppitz, Pollney, Seidel, and Takahashi}}]{Alcubierre02a}
\bibinfo{author}{\bibfnamefont{M.}~\bibnamefont{Alcubierre}},
  \bibinfo{author}{\bibfnamefont{B.}~\bibnamefont{Br\"ugmann}},
  \bibinfo{author}{\bibfnamefont{P.}~\bibnamefont{Diener}},
  \bibinfo{author}{\bibfnamefont{M.}~\bibnamefont{Koppitz}},
  \bibinfo{author}{\bibfnamefont{D.}~\bibnamefont{Pollney}},
  \bibinfo{author}{\bibfnamefont{E.}~\bibnamefont{Seidel}}, \bibnamefont{and}
  \bibinfo{author}{\bibfnamefont{R.}~\bibnamefont{Takahashi}},
  \bibinfo{journal}{Phys. Rev. D} \textbf{\bibinfo{volume}{67}},
  \bibinfo{pages}{084023} (\bibinfo{year}{2003}), \eprint{gr-qc/0206072}.

\bibitem[{\citenamefont{Br{\"u}gmann et~al.}(2004)\citenamefont{Br{\"u}gmann,
  Tichy, and Jansen}}]{Bruegmann03}
\bibinfo{author}{\bibfnamefont{B.}~\bibnamefont{Br{\"u}gmann}},
  \bibinfo{author}{\bibfnamefont{W.}~\bibnamefont{Tichy}}, \bibnamefont{and}
  \bibinfo{author}{\bibfnamefont{N.}~\bibnamefont{Jansen}},
  \bibinfo{journal}{Phys. Rev. Lett.} \textbf{\bibinfo{volume}{92}},
  \bibinfo{pages}{211101} (\bibinfo{year}{2004}),
  \bibinfo{note}{gr-qc/0312112}.

\bibitem[{\citenamefont{Reimann and Br\"ugmann}(2004)}]{mypaper1}
\bibinfo{author}{\bibfnamefont{B.}~\bibnamefont{Reimann}} \bibnamefont{and}
  \bibinfo{author}{\bibfnamefont{B.}~\bibnamefont{Br\"ugmann}},
  \bibinfo{journal}{Phys. Rev. D} \textbf{\bibinfo{volume}{69}},
  \bibinfo{pages}{044006} (\bibinfo{year}{2004}), \eprint{gr-qc/0307036}.

\bibitem[{\citenamefont{Reimann}(2003)}]{mythesis}
\bibinfo{author}{\bibfnamefont{B.}~\bibnamefont{Reimann}}, Master's thesis,
  \bibinfo{school}{Universit\"at Potsdam}, \bibinfo{address}{Germany}
  (\bibinfo{year}{2003}).

\bibitem[{\citenamefont{Smarr and York}(1978)}]{Smarr78b}
\bibinfo{author}{\bibfnamefont{L.}~\bibnamefont{Smarr}} \bibnamefont{and}
  \bibinfo{author}{\bibfnamefont{J.}~\bibnamefont{York}},
  \bibinfo{journal}{Phys. Rev. D} \textbf{\bibinfo{volume}{17}},
  \bibinfo{pages}{2529 to 2551} (\bibinfo{year}{1978}).

\bibitem[{\citenamefont{Anninos
  et~al.}(1995{\natexlab{b}})\citenamefont{Anninos, Daues, Mass{\'o}, Seidel,
  and Suen}}]{Anninos95_2}
\bibinfo{author}{\bibfnamefont{P.}~\bibnamefont{Anninos}},
  \bibinfo{author}{\bibfnamefont{G.}~\bibnamefont{Daues}},
  \bibinfo{author}{\bibfnamefont{J.}~\bibnamefont{Mass{\'o}}},
  \bibinfo{author}{\bibfnamefont{E.}~\bibnamefont{Seidel}}, \bibnamefont{and}
  \bibinfo{author}{\bibfnamefont{W.-M.} \bibnamefont{Suen}},
  \bibinfo{journal}{Phys. Rev. D} \textbf{\bibinfo{volume}{51}},
  \bibinfo{pages}{5562 to 5578} (\bibinfo{year}{1995}{\natexlab{b}}),
  \bibinfo{note}{gr-qc/9412069}.

\bibitem[{\citenamefont{Einstein and Rosen}(1935)}]{Einstein35}
\bibinfo{author}{\bibfnamefont{A.}~\bibnamefont{Einstein}} \bibnamefont{and}
  \bibinfo{author}{\bibfnamefont{N.}~\bibnamefont{Rosen}},
  \bibinfo{journal}{Phys. Rev} \textbf{\bibinfo{volume}{48}},
  \bibinfo{pages}{73 to 77} (\bibinfo{year}{1935}).

\bibitem[{\citenamefont{Geyer and Herold}(1995)}]{Geyer95}
\bibinfo{author}{\bibfnamefont{A.}~\bibnamefont{Geyer}} \bibnamefont{and}
  \bibinfo{author}{\bibfnamefont{H.}~\bibnamefont{Herold}},
  \bibinfo{journal}{Phys. Rev. D} \textbf{\bibinfo{volume}{52}},
  \bibinfo{pages}{6182 to 6185} (\bibinfo{year}{1995}).

\bibitem[{\citenamefont{Thornburg}(1993)}]{Thornburg93}
\bibinfo{author}{\bibfnamefont{J.}~\bibnamefont{Thornburg}}, Ph.D. thesis,
  \bibinfo{school}{University of British Columbia},
  \bibinfo{address}{Vancouver, Canada} (\bibinfo{year}{1993}).

\bibitem[{\citenamefont{Duncan}(1985)}]{Duncan85}
\bibinfo{author}{\bibfnamefont{M.}~\bibnamefont{Duncan}},
  \bibinfo{journal}{Phys. Rev. D} \textbf{\bibinfo{volume}{31}},
  \bibinfo{pages}{1267 to 1272} (\bibinfo{year}{1985}).

\bibitem[{\citenamefont{Alcubierre
  et~al.}(2001{\natexlab{b}})\citenamefont{Alcubierre, Brandt, Br\"ugmann,
  Holz, Seidel, Takahashi, and Thornburg}}]{Alcubierre99a}
\bibinfo{author}{\bibfnamefont{M.}~\bibnamefont{Alcubierre}},
  \bibinfo{author}{\bibfnamefont{S.}~\bibnamefont{Brandt}},
  \bibinfo{author}{\bibfnamefont{B.}~\bibnamefont{Br\"ugmann}},
  \bibinfo{author}{\bibfnamefont{D.}~\bibnamefont{Holz}},
  \bibinfo{author}{\bibfnamefont{E.}~\bibnamefont{Seidel}},
  \bibinfo{author}{\bibfnamefont{R.}~\bibnamefont{Takahashi}},
  \bibnamefont{and}
  \bibinfo{author}{\bibfnamefont{J.}~\bibnamefont{Thornburg}},
  \bibinfo{journal}{Int. J. Mod. Phys. D} \textbf{\bibinfo{volume}{10}},
  \bibinfo{pages}{273} (\bibinfo{year}{2001}{\natexlab{b}}),
  \bibinfo{note}{gr-qc/9908012}.

\bibitem[{\citenamefont{Brandt and Seidel}(1995)}]{Brandt94}
\bibinfo{author}{\bibfnamefont{S.}~\bibnamefont{Brandt}} \bibnamefont{and}
  \bibinfo{author}{\bibfnamefont{E.}~\bibnamefont{Seidel}},
  \bibinfo{journal}{Phys. Rev. D} \textbf{\bibinfo{volume}{52}},
  \bibinfo{pages}{856 to 869} (\bibinfo{year}{1995}),
  \bibinfo{note}{gr-qc/9412072}.

\bibitem[{\citenamefont{Choptuik}(1986)}]{Choptuik86}
\bibinfo{author}{\bibfnamefont{M.}~\bibnamefont{Choptuik}}, Ph.D. thesis,
  \bibinfo{school}{University of British Columbia, Vancouver, Canada}
  (\bibinfo{year}{1986}).

\bibitem[{\citenamefont{Murchadha}()}]{pc_Murchadha}
\bibinfo{author}{\bibfnamefont{N.~O.} \bibnamefont{Murchadha}},
  \bibinfo{howpublished}{private communication}.

\bibitem[{\citenamefont{Reimann}(2004)}]{mypaper3}
\bibinfo{author}{\bibfnamefont{B.}~\bibnamefont{Reimann}}
  (\bibinfo{year}{2004}), \eprint{gr-qc/0404118}.

\bibitem[{\citenamefont{Camarda}(1998)}]{Camarda97}
\bibinfo{author}{\bibfnamefont{K.}~\bibnamefont{Camarda}}, Ph.D. thesis,
  \bibinfo{school}{University of Illinois at Urbana-Champaign},
  \bibinfo{address}{Urbana, Illinois} (\bibinfo{year}{1998}).

\bibitem[{\citenamefont{Malec and Murchadha}(2003)}]{Malec03}
\bibinfo{author}{\bibfnamefont{E.}~\bibnamefont{Malec}} \bibnamefont{and}
  \bibinfo{author}{\bibfnamefont{N.~O.} \bibnamefont{Murchadha}},
  \bibinfo{journal}{Phys. Rev. D} \textbf{\bibinfo{volume}{68}},
  \bibinfo{pages}{124019} (\bibinfo{year}{2003}), \eprint{gr-qc/0307046 and
  gr-qc/0307047}.

\bibitem[{\citenamefont{Alcubierre
  et~al.}(2001{\natexlab{c}})\citenamefont{Alcubierre, Br\"ugmann, Pollney,
  Seidel, and Takahashi}}]{Alcubierre01}
\bibinfo{author}{\bibfnamefont{M.}~\bibnamefont{Alcubierre}},
  \bibinfo{author}{\bibfnamefont{B.}~\bibnamefont{Br\"ugmann}},
  \bibinfo{author}{\bibfnamefont{D.}~\bibnamefont{Pollney}},
  \bibinfo{author}{\bibfnamefont{E.}~\bibnamefont{Seidel}}, \bibnamefont{and}
  \bibinfo{author}{\bibfnamefont{R.}~\bibnamefont{Takahashi}},
  \bibinfo{journal}{Phys. Rev. D} \textbf{\bibinfo{volume}{64}},
  \bibinfo{pages}{061501} (\bibinfo{year}{2001}{\natexlab{c}}),
  \bibinfo{note}{gr-qc/0104020}.

\bibitem[{\citenamefont{Alcubierre and Br\"{u}gmann}(2001)}]{Alcubierre00c}
\bibinfo{author}{\bibfnamefont{M.}~\bibnamefont{Alcubierre}} \bibnamefont{and}
  \bibinfo{author}{\bibfnamefont{B.}~\bibnamefont{Br\"{u}gmann}},
  \bibinfo{journal}{Phys. Rev. D} \textbf{\bibinfo{volume}{63}},
  \bibinfo{pages}{104006} (\bibinfo{year}{2001}),
  \bibinfo{note}{gr-qc/0008067}.

\bibitem[{\citenamefont{Gentle et~al.}(2001)\citenamefont{Gentle, Holz,
  Kheyfets, Laguna, Miller, and Shoemaker}}]{Gentle2001}
\bibinfo{author}{\bibfnamefont{A.}~\bibnamefont{Gentle}},
  \bibinfo{author}{\bibfnamefont{D.}~\bibnamefont{Holz}},
  \bibinfo{author}{\bibfnamefont{A.}~\bibnamefont{Kheyfets}},
  \bibinfo{author}{\bibfnamefont{P.}~\bibnamefont{Laguna}},
  \bibinfo{author}{\bibfnamefont{W.}~\bibnamefont{Miller}}, \bibnamefont{and}
  \bibinfo{author}{\bibfnamefont{D.}~\bibnamefont{Shoemaker}},
  \bibinfo{journal}{Phys. Rev. D} \textbf{\bibinfo{volume}{63}},
  \bibinfo{pages}{064024} (\bibinfo{year}{2001}), \bibinfo{note}{0005113}.

\bibitem[{\citenamefont{Bronstein et~al.}(1997)\citenamefont{Bronstein,
  Semendjajew, Musiol, and M\"uhlig}}]{Bronstein97}
\bibinfo{author}{\bibfnamefont{I.~A.} \bibnamefont{Bronstein}},
  \bibinfo{author}{\bibfnamefont{K.~A.} \bibnamefont{Semendjajew}},
  \bibinfo{author}{\bibfnamefont{G.}~\bibnamefont{Musiol}}, \bibnamefont{and}
  \bibinfo{author}{\bibfnamefont{H.}~\bibnamefont{M\"uhlig}},
  \emph{\bibinfo{title}{Taschenbuch der Mathematik}}
  (\bibinfo{publisher}{Verlag Harry Deutsch}, \bibinfo{address}{Thun und
  Frankfurt am Main, Germany}, \bibinfo{year}{1997}).

\bibitem[{\citenamefont{Wald}(1984)}]{Wald84}
\bibinfo{author}{\bibfnamefont{R.~M.} \bibnamefont{Wald}},
  \emph{\bibinfo{title}{General Relativity}} (\bibinfo{publisher}{The
  University of Chicago Press}, \bibinfo{address}{Chicago},
  \bibinfo{year}{1984}).

\end{thebibliography}


\end{document}